\documentclass[desactivate]{aa}  

\usepackage{natbib}
\bibpunct{(}{)}{;}{a}{}{,}

\usepackage{graphicx}
\usepackage{revsymb}	
\usepackage{amsmath}
\usepackage{txfonts}
\usepackage{amssymb}
\usepackage{geometry} 
\usepackage[parfill]{parskip} 
\usepackage{slantsc}
\usepackage{array}
\usepackage{amsfonts}
\usepackage{siunitx}
	
\usepackage{epstopdf}	
\usepackage{epsfig}	

\usepackage{url}

\usepackage[colorlinks=true,linkcolor=black, urlcolor=black, citecolor=black]{hyperref}

\usepackage{xcolor}

\usepackage{xfrac}

\usepackage{sidecap}
\usepackage[font=small, labelfont=bf]{caption}
\usepackage{subcaption}

\usepackage{multirow}
\usepackage{booktabs}
\usepackage{threeparttable}

\usepackage[section]{placeins}

\begin{document}
   \title{Asteroseismic modelling of main-sequence solar-like stars and \textit{Kepler} exoplanet host stars with the FICO procedure }
      \subtitle{I. Catalogue of fundamental stellar properties}
\author{J. B\'{e}trisey\inst{1} \and D.~R. Reese\inst{2} \and C. Pezzotti\inst{3,4} \and M.~J. Goupil\inst{2} \and M.~S. Cunha\inst{5}}
\institute{Department of Physics and Astronomy, Uppsala University, Box 516, SE-751 20 Uppsala, Sweden\\email: 	\texttt{jerome.betrisey@physics.uu.se}
\and LIRA, Observatoire de Paris, Université PSL, Sorbonne Université, Université Paris Cité, CY Cergy Paris Université, CNRS, 92190 Meudon, France
\and STAR Institute, Université de Liège, Liège, Belgium
\and Istituto Nazionale di Astrofisica – Osservatorio Astronomico di Roma, Via Frascati 33, I-00040 Monteporzio Catone, Italy
\and Instituto de Astrofísica e Ciências do Espaço, Universidade do Porto, CAUP, Rua das Estrelas, P-4150-762 Porto, Portugal}
\date{\today}

\abstract
{Asteroseismology has revolutionised our ability to characterise solar-like stars. Since asteroseismic modelling is a key factor in upcoming space-based missions such as PLATO, it becomes increasingly relevant to develop robust and scalable modelling techniques for the precise and accurate characterisation of main-sequence and exoplanet host stars in the PLATO era.}
{We present detailed asteroseismic modelling of 95 main-sequence solar-like stars and \textit{Kepler} exoplanet host stars using the Forward and Inverse COmbination (FICO) procedure, a three-step method that combines forward and inverse techniques that enables precise inference of fundamental stellar parameters such as mass, radius, age, and mean density. These results are then discussed in the framework of the PLATO mission.}
{We applied the FICO procedure to a catalogue of nearly one hundred stars with high-quality asteroseismic and classical observations, and compared its results against literature values. We also compared its performance with direct frequency fitting using semi-empirical surface corrections.}
{The FICO procedure achieved statistical precisions of 2.3\%, 0.82\%, 6.9\%, and 0.49\% in mass, radius, age, and mean density, respectively on average, well within PLATO quality requirements. We reconfirmed that surface-independent methods more effectively mitigate biases inherent to semi-empirical surface corrections, particularly for stars more massive than $1.15 M_\odot$ or above 6050 K. Two regimes were identified: near-solar conditions, where both approaches perform similarly, and higher-mass stars, where surface-independent methods consistently outperform direct fitting methods. While our results are consistent with literature values, we observed age biases ($\sim 11.5\%$ on average for the \textit{Kepler} LEGACY sample) that are comparable to the PLATO accuracy requirement of 10\% for a Sun-like star, and therefore not negligible in that context.}
{The FICO procedure provides a robust framework for high-precision stellar characterisation in the PLATO era. Its hybrid architecture effectively addresses surface effects, making it a promising tool for the accurate determination of exoplanet host-star properties. Our findings also highlight the importance of carefully selecting and validating the physical assumptions embedded in stellar models, particularly in the context of next-generation space missions such as PLATO.}

\keywords{Asteroseismology -- Catalogs -- Stars: low-mass -- Stars: fundamental parameters -- planets and satellites: fundamental parameters -- Stars: oscillations}

\maketitle

\defcitealias{Betrisey2023_AMS_surf}{JB23}

\section{Introduction}
Space-based photometric missions, such as CoRoT \citep{Baglin2009}, \emph{Kepler} \citep{Borucki2010}, K2 \citep{Howell2014}, and TESS \citep{Ricker2015}, ushered a new era in stellar astrophysics. At the heart of this revolution lies asteroseismology, a powerful technique that allows us to probe the internal structures of stars with unprecedented precision and accuracy. These advances have dramatically enhanced our ability to determine fundamental stellar properties, with far-reaching consequences across multiple domains of astrophysics, such as galactic archaeology and exoplanetology \citep[see e.g.][for reviews and textbooks]{Chaplin&Miglio2013,Garcia&Ballot2019,Aerts2021}. In particular, accurate knowledge of the host-star is essential for precise characterization of planetary systems. Throughout their evolution, stars and planets interact through a multitude of mechanisms \citep[e.g.][]{Lanza2015,Vidotto2020}, which are closely connected to the structural and rotational properties of the host star \citep[e.g.][]{Privitera2016_spi1,Privitera2016_RG,Privitera2016_spi2,Rao2018,Barker2020,Pezzotti2021,Pezzotti2022,Pezzotti2026,Ahuir2021,
Betrisey2022,Fellay2023}. Therefore, obtaining an advanced description of the stellar evolutionary history and present properties is crucial for accurately depicting the current state of planetary systems and predicting their future and past evolution. Looking ahead, asteroseismic modeling will play a key role to meet the mission requirements of upcoming PLATO \citep{Rauer2025} and CubeSpec \citep{Bowman2022} missions, and it will be a central component of the candidate mission HAYDN \citep{Miglio2021}, dedicated to asteroseismology.

In addition to its remarkable achievements, asteroseismology has uncovered significant discrepancies in the physics of theoretical stellar models, which have the potential to bias stellar characterization at the precision level demanded by space-based photometry missions. The inaccuracies stem from uncertainties in micro- and macrophysical inputs \citep[e.g.][]{Lebreton2014,Nsamba2018,Buldgen2019f,Farnir2020,Betrisey2022}, from the treatment of near-surface layers \citep[e.g.][]{Ball&Gizon2017,Nsamba2018,Jorgensen2020,Jorgensen2021,Cunha2021,Betrisey2023_AMS_surf}, and from magnetic activity effects \citep[e.g.][]{Creevey2011,PerezHernandez2019,Thomas2021,Betrisey2024_MA_Sun,
Betrisey2025_MA_Inv,Betrisey2025_MA_Damping}. With the PLATO mission poised for launch in late 2026, the community is intensifying efforts to refine and benchmark the performance of optimization algorithms used to infer fundamental stellar parameters, such as mass, radius, age, and mean density, from oscillation spectra \citep[see e.g.][and references therein]{Silva-Aguirre2015,Silva-Aguirre2017,Cunha2021}. Over the years, a diverse suite of methodologies has emerged, including Bayesian inference \citep{Silva-Aguirre2015,Silva-Aguirre2017,AguirreBorsen-Koch2022}, Markov chain Monte Carlo methods \citep[MCMC;][]{Bazot2008,Gruberbauer2013,Rendle2019,Jiang&Gizon2021}, genetic algorithms \citep{Metcalfe&Charbonneau2003,Charpinet2005,Metcalfe2009,Metcalfe2014,Metcalfe2023}, Levenberg-Marquardt algorithms \citep[see e.g.][]{Frandsen2002,Teixeira2003,Miglio&Montalban2005}, and machine-learning methods \citep{Bellinger2016,Hon2020,Guo&Jiang2023}. Interestingly, simulated annealing methods \citep[e.g.][]{Kirkpatrick1983,Kirkpatrick1984,Press1986} remain largely unexplored in this context.

To date, the pool of main-sequence solar-like stars with high-quality seismic data comprises roughly a hundred targets, thanks largely to the legacy of the \textit{Kepler} mission \citep[e.g.][]{Davies2016,Lund2017}. Among those, about a quarter are confirmed exoplanet host stars \citep[see e.g.][and references therein]{Silva-Aguirre2015}. In this study, we derive the fundamental stellar parameters of these prime stars using the Forward and Inverse COmbination (FICO) procedure \citep{Betrisey2022,Betrisey2023_AMS_surf,Betrisey2024_AMS_quality,Betrisey2024_phd}, a modelling framework that integrates both forward and inverse techniques, and interpret these results in the context of the PLATO mission. Our analysis builds directly on the methodological foundations established two years ago in \citet{Betrisey2023_AMS_surf}, which defined the core assumptions and algorithmic choices of the FICO approach. That earlier work already addressed several key methodological aspects. Specifically, it examined: 1) the accuracy of FICO using synthetic models constructed by patching 3D hydrodynamical simulations to 1D stellar structures, followed by non‑adiabatic frequency computations, as well as tests on well‑established calibrators from the literature, including the binary system 16 Cygni; 2) the treatment of surface effects and their impact on the inferred parameters; 3) the choice of seismic and non‑seismic constraints, including an in‑depth analysis of their influence under different modelling assumptions; and 4) the handling of unstable or non‑convergent solutions. Building on this foundation, the present work constitutes a catalogue‑type follow‑up study and leverages the larger statistical sample to explore complementary aspects that could not be robustly addressed in \citet{Betrisey2023_AMS_surf}. Specifically, we focus on three main aspects: 1) comparison between the surface-independent FICO procedure and an older surface-dependent approach, providing feedback for PLATO's internal ranking of modelling techniques for stellar parameter estimation; 2) discussion of the intrinsic statistical precision and whether it meets PLATO's quality requirements; and 3) comparison with literature values, identifying potential outliers.

We also note that our stellar models are computed assuming the AGSS09 solar abundance mixture \citep{Asplund2009}. This choice is particularly relevant in the context of asteroseismic catalogues, as widely-used stellar databases \citep[e.g.][]{Silva-Aguirre2015,Silva-Aguirre2017} often rely on the higher‑metallicity abundances of \citet{Grevesse&Sauval1998}. As a result, the catalogue presented here provides a complementary resource that will support assessments of systematic uncertainties linked to the choice of reference composition within the PLATO framework.

In Sect.~\ref{sec:presentation_stellar_catalogue}, we introduce the stellar catalogue and describe the observational data selected for our analysis. Sect.~\ref{sec:modelling_strategy} details the grid of stellar models employed in the minimisation process, along with the overall modelling strategy. In Sect.~\ref{sec:results}, we present the fundamental stellar parameters derived using the FICO modelling procedure. In Sect.~\ref{sec:discussion}, we compare our results with those obtained from direct fitting of individual oscillation frequencies, as well as with values reported in the literature. We also discuss the precision of our methodology. Finally, Sect.~\ref{sec:conclusions} summarises the main findings of this first paper.

\section{Presentation of the stellar catalogue}
\label{sec:presentation_stellar_catalogue}
Our stellar catalogue combines the LEGACY \citep{Lund2017,Silva-Aguirre2017} and \citet{Davies2016} samples, which together represent the most comprehensive high-quality seismic datasets of main-sequence solar-like stars available prior to the launch of the PLATO mission \citep{Rauer2025}.

\subsection{Seismic data}
\label{sec:seismic_data}
The LEGACY sample includes 66 \textit{Kepler} targets, while the \citet{Davies2016} sample comprises 35 \textit{Kepler} Objects of Interest (KOIs). These two datasets currently hold the highest-quality seismic observations in this stellar regime: the stars were observed at high cadence over long baselines, enabling high frequency resolution and correspondingly small formal uncertainties on individual mode frequencies. While PLATO will not systematically match the full \textit{Kepler} observing baseline for every target, its core programme nonetheless includes long‑duration pointings of about two years per field for the brightest stars. These are precisely the targets for which peak‑bagging of solar‑like oscillations is both feasible and intended \citep{Rauer2025}. Since the frequency resolution scales inversely with the observing duration $T_{\rm obs}$ \citep[see e.g.][]{Appourchaux1998,Appourchaux2014}, a two‑year PLATO time series yields a frequency resolution approximately a factor of two coarser than that obtained from the four‑year \textit{Kepler} baseline. The precision of fitted mode frequencies, however, scales more weakly with observing time, following a $T_{\rm obs}^{-1/2}$ dependence for resolved, stochastically excited solar‑like oscillations. Consequently, reducing the observing baseline by a factor of two degrades the mode‑frequency precision by only a factor of $\sqrt{2}$, and the expected performance for the best PLATO dwarfs is correspondingly `LEGACY-like'. In particular, \citet{Goupil2024} predict that the PLATO core programme will increase the population of prime main‑sequence targets with LEGACY‑level seismic characterisation to approximately 800--1000 stars.

Two stars (KIC7199397 and KIC8684730) from the \citet{Davies2016} sample were excluded due to the presence of mixed modes, hinting that these stars might in fact be subgiant stars, which fall outside the scope of our analysis. Additionally, four targets (KIC3632418, KIC9414417, KIC9955598, and KIC10963065) are shared between both samples; however, since their seismic data were independently derived by separate teams, we provide two distinct sets of fundamental parameters for each of these stars. This brings the total number of unique observed targets to 95.

The LEGACY sample also includes a 67th target, representing solar data artificially degraded to match the quality of \textit{Kepler} observations. Originally introduced as a control target, this dataset was known to the modelling teams of the \citet{Silva-Aguirre2017} paper and used to benchmark the reliability of their algorithms. Although its true nature has now been revealed, we retain this control in our analysis by considering two versions: one using the unperturbed non-seismic solar constraints, and another using the perturbed non-seismic constraints as defined in \citet{Silva-Aguirre2017}.

As previously highlighted by \citet{Betrisey2022} for Kepler-93 and more extensively by \citet{Betrisey2024_phd} across the full sample, 19 out of the 33 targets from \citet{Davies2016} are likely (and surely for Kepler-93) to exhibit inconsistent mode identifications, where radial orders are shifted by a constant factor ranging from -1 to +2 depending on the target. It is important to note that mode identification is a theoretical quantity, assigned a posteriori to the observed individual oscillation frequencies. While inconsistencies in mode identification do not compromise the quality of the seismic data itself, they introduce challenges in the determination of fundamental stellar parameters, as mode identification is employed to improve the convergence of the minimisation algorithms. Typically, incorrect mode identification leads to minimisation failure, characterized by a specific mismatch between the optimal theoretical frequencies and the observed échelle diagram. In rare cases, however, we observed that the algorithm may converge to an incorrect solution, which can be identified through manual inspection by an experienced modeller. Such cases pose a serious concern for automated pipelines, such as in the PLATO stellar modelling framework, where manual verification is not feasible. Several studies \citep[e.g.][]{White2011,Roxburgh2016,Betrisey2024_phd} suggested that inconsistencies in mode identification could be inferred from the $\varepsilon_{nl}$ phase values, which are expected to lie systematically between 1 and 2 \citep[using the phase definition of][]{Roxburgh2016} for main-sequence solar-like stars. Our detailed modelling of the \citet{Davies2016} targets confirms this hypothesis for this sample. A solution to mitigate this issue in a pipeline would be to validate mode identification using $\varepsilon_{nl}$ phases, a method already employed by some observers \citep[e.g.][]{Appourchaux2012}.  For completeness, we note that Table~4.5 of \citet{Betrisey2024_phd} provides the correction factor to obtain the accurate mode identification of the 19 problematic targets.

\subsection{Classical constraints}
\label{sec:classical_constraints}
Classical constraints refer to stellar observables that are independent of the individual oscillation frequencies or any derived combinations thereof. In our analysis, the primary classical constraints are the effective temperature, metallicity, and absolute luminosity. For targets where the luminosity could not be reliably determined, we substituted it with the frequency of maximum oscillation power $\nu_{\rm max}$ as an alternative constraint.

Spectroscopic parameters were primarily sourced from \citet{Furlan2018}. For targets not included in that catalogue, we adopted the values provided by \citet{Pinsonneault2012}. The $\nu_{\rm max}$ frequencies are taken from \citet{Lund2017} and \citet{Davies2016}. The absolute luminosity was computed with
\begin{equation}
-2.5\log\left(\frac{L}{L_\odot}\right) = m_\lambda + BC_\lambda -5\log d + 5 - A_\lambda -M_{\mathrm{bol},\odot}\, .
\label{eq:absolute_luminosity}
\end{equation}
The wavelength is denoted by $\lambda$. We used the 2MASS $K_s$- and $H$-bands for our analysis. The $K_s$-band was prioritised, with the $H$-band serving as a fallback in cases where the former measurement was deemed unreliable. For the solar bolometric correction, we adopted a value of $M_{\mathrm{bol},\odot} = 4.75$ and the apparent magnitude is represented by $m_\lambda$. The bolometric correction $BC_\lambda$ was inferred following \citet{Casagrande2014,Casagrande2018}, and the extinction $A_\lambda$ was computed using the dust map from \citet{Green2018}. We tested several approaches to estimate the distance $d$ in pc from \textit{Gaia} EDR3 data \citep{Gaia2021}, either inverting the \textit{Gaia} parallax including the correction from \citet{Lindegren2021} or using the distances provided by \citet{Bailer-Jones2021}. Both methods yielded consistent results, and we ultimately adopted the \citet{Bailer-Jones2021} distances for the computation of the absolute luminosity.

\section{Modelling strategy}
\label{sec:modelling_strategy}

\subsection{The Spelaion grid}
\label{sec:spelaion_grid}
The Spelaion grid, first introduced in \citet{Betrisey2023_AMS_surf}, is a comprehensive, high-resolution set of standard non-rotating stellar models originally designed to span the main-sequence evolution of solar-like stars with masses between 0.80 and 1.60 $M_\odot$. In this work, we present a substantial expansion of the grid, increasing its scope with the addition of approximately 10,000 new evolutionary tracks, bringing the total to over 35,000 tracks. This upgraded version now encompasses solar-like stars from 0.70 to 1.66 $M_\odot$, and includes a broader range of initial chemical compositions, thereby enhancing its applicability to diverse stellar populations. A detailed overview of the mesh architecture and statistical properties of the extended grid is provided in Appendix~\ref{sec:app:upgraded_spelaion_grid}.

To ensure consistency, the physical assumptions remain aligned with those in \citet{Betrisey2023_AMS_surf}. Notably, the grid is constructed using the low-metallicity abundance tables from \citet{Asplund2009}, offering a valuable complement to other widely used grids \citep[e.g.][and references therein]{Silva-Aguirre2015,Bellinger2016,Silva-Aguirre2017,Creevey2017,Metcalfe2023}, which typically rely on the high-metallicity tables from \citet{Grevesse&Sauval1998}.

\subsection{The FICO modelling procedure}
\label{sec:FICO_modelling_procedure}
Although mixing-length theory remains a widely adopted framework for approximating stellar convection (including in our own modelling), it is fundamentally limited by its simplicity \citep[see e.g.][for textbooks]{Maeder2009,Kippenhahn2012}. This approach reduces the inherently complex, three-dimensional, and non-local nature of turbulent convection to a one-dimensional, localised approximation. Such a simplification proves particularly inadequate in the outermost layers of stars \citep[e.g.][]{Nordlund&Stein1997,Stein&Nordlund1998,Tanner2013}, where the timescale of convection and the dynamical timescale become comparable. Although this shortcoming has a limited impact on broader stellar modelling, it becomes critically important in asteroseismology, as the near-surface layers are the birthplace of acoustic oscillations. Consequently, the oversimplified treatment of convection introduces systematic errors in the predicted oscillation frequencies \citep[e.g.][]{Kjeldsen2008,Ball&Gizon2014,Sonoi2015}.

To mitigate these surface-related discrepancies, we employed the FICO procedure \citep{Betrisey2022,Betrisey2023_AMS_surf,Betrisey2024_AMS_quality,Betrisey2024_phd}, a three-step modelling strategy that integrates both forward and inverse techniques to circumvent these effects and refine the seismic characterisation of solar-like stars. This methodology builds upon pioneering studies by \citet{Reese2012} and \citet{Buldgen2019f}, and unfolds in three key stages. The initial step involves constructing a reference model by fitting individual oscillation frequencies, supplemented with a semi-empirical correction to account for surface effects. Although this model may carry slight biases in its global parameters, its mean density is sufficiently close to the true value to serve as a reliable foundation for subsequent analysis. The second stage applies a mean density inversion, leveraging the non-linear extension formalism of \citet{Reese2012} to extract a quasi-model-independent correction. Finally, the third step employs frequency separation ratios, carefully designed combinations of individual frequencies that are inherently almost insensitive to surface layers \citep{Roxburgh&Vorontsov2003,Oti2005}. While these ratios sacrifice direct information about the mean density, this loss is compensated by the inversion step. We refer the reader to \citet{Buldgen2022c} for a comprehensive overview of seismic inversion techniques and their successful application to a range of asteroseismic targets \citep[see e.g.][for a non-exhaustive selection]{DiMauro2004_theo,Reese2012,Buldgen2015a,Buldgen2015b,Buldgen2016b,Buldgen2016c,Buldgen2017c,Buldgen2019b,Buldgen2019f,
Buldgen2022b,Kosovichev&Kitiashvili2020,Salmon2021,Betrisey&Buldgen2022,Betrisey2022,Betrisey2023_rot,
Betrisey2023_AMS_surf,Betrisey2024_AMS_quality,Betrisey2024_MA_Sun,Betrisey2025_MA_Inv,Betrisey2025_MA_Damping}. 

We note that mean‑density inversions are not insensitive to surface effects, since they rely on pressure modes whose eigenfunctions exhibit large amplitudes in the near‑surface layers. Nevertheless, the impact of surface effects on these inversions is reduced compared to other mean‑density indicators (such as the large frequency separation), owing to a weaker near‑surface amplitude of the inversion kernel \citep[see e.g.][]{Oti2005,Reese2012}. The effective mitigation of surface effects arises instead from the use of frequency separation ratios \citep{Roxburgh&Vorontsov2003,Oti2005}. While the inverted mean density retains some sensitivity to surface effects, this sensitivity affects only a single constraint, rather than all seismic constraints as in a direct fit of individual frequencies. The purpose of including the inverted mean density is to reintroduce the mean‑density information that is otherwise lost when using separation ratios alone. When this constraint is included, we explicitly account for its residual sensitivity to surface effects following \citet[][]{Betrisey2023_AMS_surf}. The associated inaccuracies are too small to bias the inferred stellar parameters, but they must be properly accounted for when estimating statistical uncertainties.

In practical terms, the modelling process unfolded as follows. Initially, we utilised the Asteroseismic Inference on a Massive Scale (AIMS) software \citep{Rendle2019} to perform a detailed fit of the observed oscillation frequencies alongside classical constraints, namely, effective temperature, metallicity, and absolute luminosity or $\nu_{\rm max}$ frequency. AIMS operates using a Markov chain Monte Carlo (MCMC) algorithm, implemented via the Python \textsc{emcee} package \citep{Foreman-Mackey2013}, to interpolate across a precomputed grid of stellar models. Notably, the interpolation is performed within the parameter space that defines the evolutionary tracks used to construct the grid. The set of parameters used for the computation of the interpolation coefficients is restricted to those optimised during the fitting process. Other stellar quantities are subsequently interpolated a posteriori, using the interpolation coefficients derived from the optimised variables. This Bayesian framework enables the derivation of posterior distributions for the optimised stellar parameters, facilitating a comprehensive exploration of the parameter space when paired with a high-resolution grid of stellar models. For our analysis, we employed the updated version of the Spelaion grid described in Sect.~\ref{sec:spelaion_grid} and in Appendix~\ref{sec:app:upgraded_spelaion_grid}. Surface effects were accounted for using the semi-empirical correction proposed by \citet{Ball&Gizon2014}.

The first optimisation with AIMS targeted the following stellar parameters: mass, age, initial hydrogen and metal mass fractions ($X_0$ and $Z_0$), and two coefficients governing the surface correction ($a_{3}$ and $a_{-1}$). For stars exceeding $1.10M_\odot$, the overshooting parameter was also included as a free parameter. Uniform, non-informative priors were adopted for all variables, except for stellar age, which was constrained to a uniform distribution between 0 and 13.8 Gyr. Observational uncertainties were modelled as Gaussian noise for the likelihood calculations. A weight of one is attributed to the classical and seismic constraints, meaning that their uncertainties are not altered (as would be with approaches overweighting classical constraints for example). This weighting is the most logical from a statistical standpoint and enables meaningful comparisons with the accuracy requirements set by the PLATO mission \citep[see detailed discussion in][]{Cunha2021}. As a aside, we note that this approach is known as the no weights approach in the context of the PLATO mission and tends to diminish the influence of classical constraints, which are fewer and generally less precise than their seismic counterparts. Each AIMS run was conducted in two phases: a preliminary burn-in phase to identify the relevant region of parameter space, followed by a production phase to sample the posterior distributions. For the initial frequency fitting, we employed 800 walkers, 8000 burn-in steps, and 3000 production steps, yielding approximately 2.4 million samples from the solution phase, following 6.4 million likelihood evaluations during burn-in. This extensive sampling was chosen to ensure robust convergence, particularly for the surface correction parameters \citep[see][]{Betrisey2023_AMS_surf}. Following this, a mean density inversion was performed to obtain a robust constraint on the stellar mean density. This constraint was then incorporated \citep[as a classical constraint; see discussion in][]{Betrisey2023_AMS_surf} into a second AIMS run, which focused on fitting frequency separation ratios rather than individual frequencies. This second optimisation produced 2.4 million production samples, following 2.4 million likelihood evaluations during burn-in.

Considering the potential interpolation issues near the end phase of the main sequence, where seismic diagnostics become increasingly degenerate \citep[see e.g.][]{White2011}, we conducted an additional validation to ensure the reliability of the final stage of the FICO procedure. This test aimed to confirm that the oscillation frequencies derived from the AIMS interpolation are physically consistent and reproducible by an evolutionary model (AIMS does not interpolate the model structure). To this end, we recomputed a stellar model using the optimal parameters obtained from AIMS, employing the same versions of the stellar evolution and pulsation codes used to generate the Spelaion grid. A successful interpolation would yield a model whose theoretical frequencies closely match those interpolated by AIMS, and both sets would reproduce the observed échelle diagram with comparable fidelity. In contrast, a mismatch between the two frequency sets, manifesting itself as discrepancies in the échelle diagram, would indicate that the interpolated solution is not physically viable. This issue typically causes an overestimation of the stellar age and necessitates a local re-optimisation to recover an unbiased stellar characterisation as described below.

Although in minority, such interpolation failures were observed in 19 cases (approximately 20\% of the sample), affecting the following targets: KIC2837475, KIC3656476, KIC4349452, KIC4914923, KIC5950854, KIC6508366, KIC6521045, KIC6603624, KIC6933899, KIC7296438, KIC7680114, KIC7871531, KIC8424992, KIC8694723, KIC9592705, KIC10068307, KIC10514430, KIC10516096, KIC11253226, KIC11772920, KIC12069449, and KIC12317678. For these stars, we performed an additional local minimisation using a Levenberg–Marquardt algorithm \citep[e.g.][]{Frandsen2002,Teixeira2003,Miglio&Montalban2005}, calculating on-the-fly evolutionary models and optimising the same set of free parameters as in AIMS. Tests conducted within the frameworks of \citet{Buldgen2019f} and \citet{Betrisey2022} showed that the Levenberg-Marquardt method might underestimate uncertainties for some targets. The issue is not related to the Hessian matrix, but rather to the intrinsic nature of the Levenberg-Marquardt algorithm. By design, this method estimates uncertainties based on local information, and depending on the shape of the parameter space, it can produce very small uncertainty values. While this is acceptable from a numerical and mathematical perspective, we are not convinced that such high precision is physically meaningful. For this reason, whenever the Levenberg-Marquardt algorithm returned smaller uncertainties than AIMS, we chose to adopt the latter uncertainties instead. We indeed consider that the AIMS uncertainties are more robust, owing to the extensive MCMC sampling around the solution, and we deemed it reasonable to assume that the local minimisation results lie sufficiently close to the AIMS solution to justify reusing these uncertainty estimates.

\section{Results}
\label{sec:results}

\subsection{The Sun as a control target}
\label{sec:the_sun}
\begin{table*}[t!]
\centering
\caption{Results of the consistency tests with degraded solar data.}
\begin{tabular}{lcccc}
\hline \hline \noalign{\vskip 0.5ex}
\multirow{2}{*}{Solar parameter} & \multicolumn{2}{c}{Fit frequencies + Ball \& Gizon (2014)} & \multicolumn{2}{c}{FICO} \\ \cmidrule[0.4pt](lr){2-3} \cmidrule[0.4pt](l){4-5}
 & Unperturbed & Perturbed & Unperturbed & Perturbed \\ 
\hline\noalign{\vskip 0.5ex}
Mass $(M_\odot)$ & $1.019\pm 0.009$ & $1.020\pm 0.009$ & $0.984\pm 0.019$ & $0.997\pm 0.018$ \\ 
Radius $(R_\odot)$ & $1.006\pm 0.003$ & $1.006\pm 0.003$ & $0.994\pm 0.007$ & $0.999\pm 0.006$ \\ 
Age (Myr) & $4811\pm 68$ & $4816\pm 68$ & $4642\pm 132$ & $4675\pm 145$ \\ 
Mean density (g/cm$^{3}$) & $1.4112\pm 0.0014$ & $1.4112\pm 0.0014$ & $1.4121\pm 0.0024$ & $1.4112\pm 0.0025$ \\ 
Initial helium abundance & $0.252\pm 0.004$ & $0.251\pm 0.004$ & $0.281\pm 0.017$ & $0.272\pm 0.017$ \\ 
Initial abundance of metals & $0.0153\pm 0.0010$ & $0.0153\pm 0.0010$ & $0.0153\pm 0.0021$ & $0.0154\pm 0.0022$ \\
Surface helium abundance & $0.225\pm 0.004$ & $0.225\pm 0.004$ & $0.251\pm 0.017$ & $0.243\pm 0.018$ \\ 
Effective temperature (K) & $5745\pm 25$ & $5745\pm 25$ & $5810\pm 41$ & $5789\pm 41$ \\ 
$\rm [Fe/H]$ (dex) & $0.008\pm 0.029$ & $0.008\pm 0.029$ & $0.022\pm 0.067$ & $0.020\pm 0.070$ \\ 
Luminosity $(L_\odot)$ & $0.994\pm 0.012$ & $0.993\pm 0.012$ & $1.014\pm 0.018$ & $1.009\pm 0.019$ \\ 
\hline
\end{tabular} 
\label{tab:consistency_tests}
\end{table*}

First, we recall that the accuracy of the FICO procedure was already investigated in \citet{Betrisey2023_AMS_surf} using both synthetic targets and well‑studied stars from the literature. Building on that earlier analysis, and motivated by the fact that \citet{Betrisey2023_AMS_surf} adopted a different solar reference dataset, we carried out an additional consistency check using the 67th target of the LEGACY sample, which corresponds to degraded solar data. Two distinct sets of classical constraints were tested:
\begin{enumerate}
\item Unperturbed set: $T_{\mathrm{eff}}=(5772\pm 100)$~K, $\rm [Fe/H]=(0.00\pm 0.10)$ dex, and $L = (1.00\pm 0.03)\, L_\odot$,
\item Perturbed set: $T_{\mathrm{eff}}=(5752\pm 77)$ K, $\rm [Fe/H]=(-0.02\pm 0.15)$ dex, and $L = (0.989\pm 0.03)\, L_\odot$ \citep[from][]{Silva-Aguirre2017}.
\end{enumerate}
It is worth noting that this dataset also carries historical relevance. Originally introduced as a control case, its identity was intentionally withheld from the modelling teams participating in the study of \citet{Silva-Aguirre2017}, allowing it to serve as a blind benchmark for evaluating the robustness of their pipelines. Although its solar nature is now known, it remains customary to include this target in re‑analysis of the LEGACY sample, ensuring continuity and comparability with previous work.

The FICO procedure is performing as expected, as shown in Table \ref{tab:consistency_tests}, which presents the solar parameters derived from both sets using two approaches: a direct fit of the individual oscillation frequencies employing the surface correction prescription of \citet{Ball&Gizon2014}, which corresponds to the first stage of the FICO procedure and a widely adopted method in the asteroseismic community, and the full FICO procedure, which incorporates surface-independent constraints. All configurations yield solar parameters with high accuracy, comfortably satisfying the PLATO mission's accuracy requirements for Sun-like stars (15\% in mass, 1--2\% in radius, and 10\% in age).

A closer inspection, however, reveals a slight underperformance in the direct frequency fitting approach, particularly in the estimation of the solar age and surface helium abundance. The expected values are 4.57 Gyr \citep{Connelly2012} and 0.2423 \citep{Asplund2021}, respectively. This bias likely stems from a combination of factors, including the physical assumptions embedded in the model grid and the treatment of surface effects. In our analysis, we adopted the abundances from \citet{Asplund2009}, which are known to be fundamentally incompatible (at the precision level achieved in modern solar modelling) with the opacities and equation of state used in solar models, a longstanding unresolved issue referred to as the solar modelling problem \citep[see e.g.][]{Basu&Antia2008,Buldgen2019e}. The results obtained via the full FICO procedure, which bypasses surface effects, are more accurate. Although the \citet{Ball&Gizon2014} prescription provides a physically-motivated correction for surface effects, it is an approximation and not an exact physical law. Consequently, when the MCMC algorithm adjusts the free parameters of the surface correction, residual inaccuracies must be absorbed by the other free parameters of the stellar model (in our case: mass, age, $Y_0$, and $Z_0$). These parameters are, moreover, strongly interdependent. For instance, modifying $Y_0$ invariably shifts the mass and age in the optimisation. As a result, even small inaccuracies in the surface prescription induce biases in these quantities. In our case, the MCMC tended to compensate for the residual surface mismatches by shifting mass, age, and $Y_0$ away from their expected solar values.

Additionally, for the surface-independent FICO results, we posit that the adopted abundances continue to exert an influence on the inferred stellar age, and that some degree of compensation may have occurred through the selection of a slightly more metal-rich solution. Indeed, both constraint sets yielded initial metals-to-hydrogen ratios of 0.0218 and 0.0217 for the unperturbed and perturbed cases, respectively, both exceeding the expected value of 0.0199 given our choice of abundances.

\subsection{Fundamental stellar properties}
\label{sec:fundamental_stellar_properties}
\begin{figure}[t!]
\centering
\includegraphics[scale=0.55]{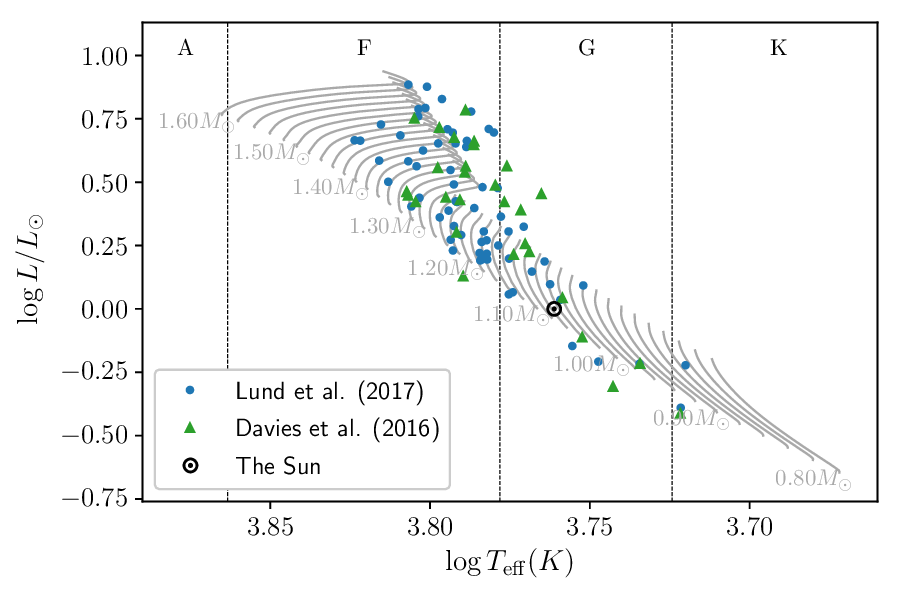} 
\caption{Hertzsprung-Russell diagram of the stellar catalogue. The Sun is shown as a reference and the letters stand for the spectral type. The grey lines represent evolutionary tracks of the Spelaion grid selected by fixing the initial chemical composition to $X_0=0.72$ and $Z_0=0.018$, and setting $\alpha_{\mathrm{ov}}=0$.}
\label{fig:hr_diagram}
\end{figure}

In Fig.~\ref{fig:hr_diagram}, we present the Hertzsprung–Russell diagram of the stellar catalogue. The LEGACY and \citet{Davies2016} samples predominantly consist of F-type stars, with only a few G-type stars included. This distribution reflects the observational bias towards brighter stars, whose oscillations are more readily detectable.

During the initial stage of the FICO procedure, meaningful physical convergence was not achieved for four targets within the Davies sample, specifically, stars KIC4143755, KIC5094751, KIC8349582, and KIC11401755. AIMS solutions were unstable as slightly modifying the observational constraints lead to incompatible solutions. Without an advanced and costly analysis of each target beyond the purpose of this study, we considered the MCMC solutions as non-physical. Consequently, these stars were excluded from the final catalogue. Without undertaking a more detailed, individualised analysis beyond the scope of this study, it remains challenging to determine the precise cause of this failure. However, several plausible explanations may account for this outcome: insufficient reliability of the seismic data (due a low signal-to-noise ratio), misidentification of one or several oscillation modes, the presence of non-standard stellar ingredients not adequately captured by the standard models employed in our grid, or grid boundary issues near the end of the main-sequence. Furthermore, the third stage of the FICO procedure failed to converge (to a meaningful physical solution) for KIC8938364 from the LEGACY sample. Given that the failure occurred only at the final step, we suspect that non-standard physical processes are the most likely cause. A comprehensive list of the derived fundamental parameters is provided in Table~\ref{tab:complete_fundamental_stellar_data}.

Furthermore, as discussed in Sect.~\ref{sec:presentation_stellar_catalogue}, four targets are shared between the two samples: KIC3632418, KIC9414417, KIC9955598, and KIC10963065. Because the light curves of these stars have been analysed independently by two different teams, we include both sets of asteroseismic characterisations in Table~\ref{tab:complete_fundamental_stellar_data} for completeness. The derived mass, radius, and age parameters for these four stars agree remarkably well, with uncertainty intervals overlapping at $1\sigma$, underscoring the consistency and robustness of our methodology.

\section{Discussion}
\label{sec:discussion}

\subsection{Treatment of surface effects}
\label{sec:treatment_SE}
Given that the initial stage of the FICO procedure involves a direct fitting of individual oscillation frequencies by assuming a semi-empirical correction for surface effects, a methodology still widely employed within the community and part of the age estimation techniques employed in the PLATO stellar modelling pipeline, we are able to compare its outcomes with those derived from the complete FICO procedure, which relies on surface-independent constraints.

Previous investigations have explored this comparison for both main-sequence and post-main-sequence stars, albeit with limited sample sizes \citep[e.g.][]{Ball&Gizon2017,Nsamba2018,Jorgensen2020,Jorgensen2021,Cunha2021,Betrisey2023_AMS_surf}. These studies consistently found that the direct fitting approach tends to overestimate stellar mass and radius, while underestimating stellar age. At the precision level targeted by PLATO, however, the differences between the two methodologies become more nuanced \citep[see e.g.][and references therein]{Betrisey2023_AMS_surf}. While both approaches yield comparable results for stars in the vicinity of the Sun, discrepancies emerge at higher stellar masses. The sample analysed by \citet{Betrisey2023_AMS_surf} was however too limited to robustly identify the transition between these regimes. 

\begin{figure}[t!]
\centering
\begin{subfigure}[b]{.41\textwidth}
  \includegraphics[width=.99\textwidth]{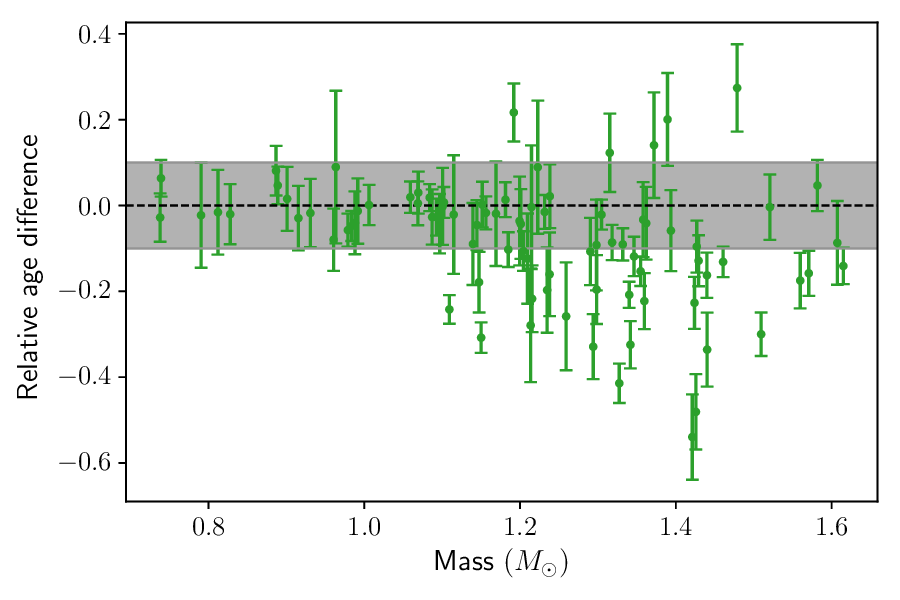}  
\end{subfigure}
\begin{subfigure}[b]{.41\textwidth}
  \includegraphics[width=.99\textwidth]{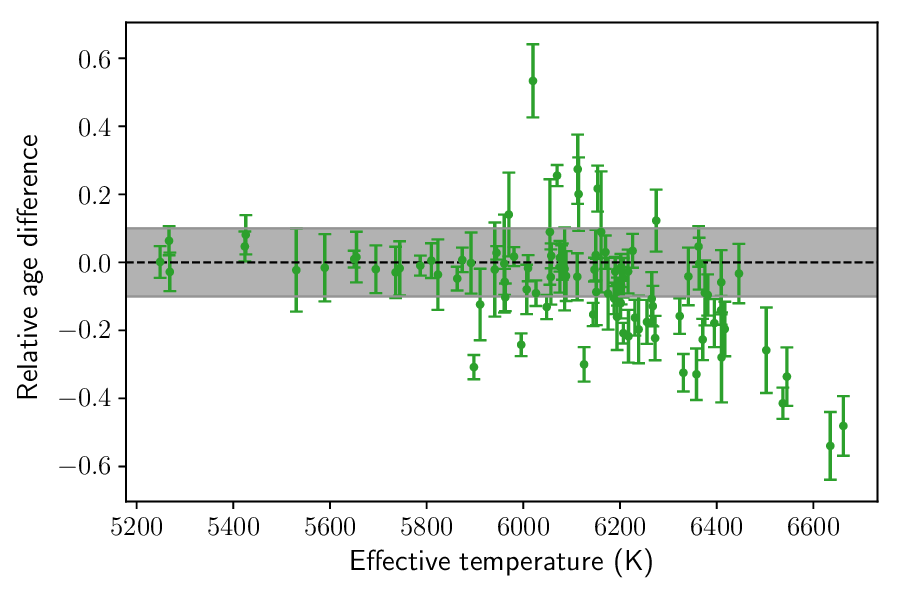}  
\end{subfigure}
\begin{subfigure}[b]{.41\textwidth}
  \includegraphics[width=.99\textwidth]{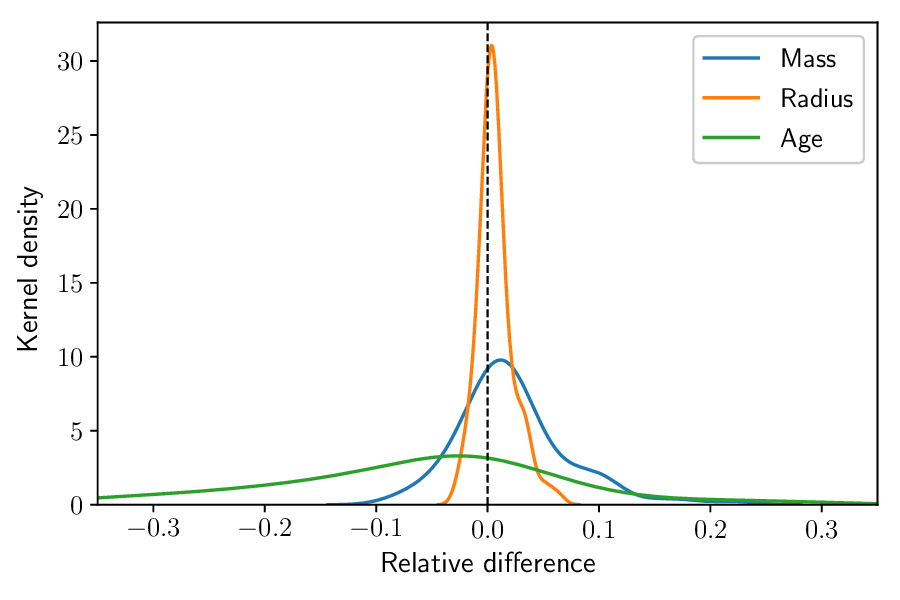}  
\end{subfigure}
\caption{\textit{Upper panels:} Relative age difference between a direct fit of the individual frequencies as a function of stellar mass and effective temperature, assuming a semi-empirical prescription for the surface effects, and the FICO procedure, based on surface-independent constraints. The grey area represents the PLATO requirement in age for a Sun-like star. \textit{Lower panel:} Kernel density estimates of the relative differences in mass (blue), radius (orange), and age (green).}
\label{fig:relative_differences_SE}
\end{figure}

Our results corroborate and extend these earlier findings, as illustrated in Fig.~\ref{fig:relative_differences_SE}. The upper panels display the relative age difference, a parameter especially sensitive to modelling choices and a key focus of PLATO, as a function of stellar mass and effective temperature. Owing to the larger statistical sample considered here, we are now able to more clearly delineate the transition between the regimes where surface‑dependent and surface‑independent approaches yield comparable or discrepant results. As discussed later in this section, this transition cannot be encapsulated by a sharp threshold in either mass or effective temperature alone. Instead, its physical origin is likely multidimensional, involving chemical composition and other stellar properties. A dedicated investigation, beyond the purpose of our study, would thus be required to fully characterise the underlying physics and to map the relevant parameter space. Within the practical framework of the PLATO mission, however, simple and robust criteria are needed to rank the reliability of the various modelling techniques. In this context, approximate thresholds in mass or effective temperature beyond which surface‑independent methods should systematically be prioritised already represent a useful and actionable outcome.

To characterise the transition, defined by whether the relative deviation falls within the tolerance bounds set by PLATO’s precision requirements (5\% in mass\footnote{The official PLATO requirement is 15\% for the stellar mass based on precision requirements for the planetary parameters, but for consistency with the stellar age, a precision of 3\% to 5\% is required.}, 2\% in radius, and 10\% in age), we employed a decision tree classifier with a single split (maximum depth = 1) using the dedicated routine from the Scikit-learn Python package \citep{Scikit-learn2011}. The classifier was trained using stellar mass as the only feature and a binary target indicating whether each data point lies within or outside the specified tolerance interval. To account for observational uncertainties, we assigned sample weights inversely proportional to the reported uncertainties, thereby reducing the influence of noisier data during the optimisation of the split. To estimate the statistical uncertainty of the transition threshold, we applied a bootstrap resampling procedure with 1000 iterations \citep[see e.g.][for an introduction to bootstrap]{Efron&Tibshirani1993}. In each iteration, a new dataset was drawn with replacement, and the threshold was re-estimated using the same weighted decision tree approach. The standard deviation of the resulting threshold distribution was adopted as the uncertainty.

Expressed in terms of stellar mass, we infer transition thresholds of $1.21\pm0.07\, M_\odot$, $1.21\pm0.06\, M_\odot$, and $1.18\pm0.06\, M_\odot$ when considering deviations in mass, radius, and age, respectively. Expressed in terms of effective temperature, the corresponding thresholds are $6198\pm153$~K, $6198\pm89$~K, and $6207\pm125$~K. In the context of the PLATO mission, these results are already highly informative. They indicate that surface‑independent modelling approaches should be systematically favoured for stars with masses exceeding approximately $1.15 M_\odot$ or with effective temperatures above roughly 6050~K.

While a dedicated study would be required to confirm this hypothesis, a plausible explanation for the observed transition lies in the onset of a persistent convective core. Although not directly linked to the surface treatment itself, the presence of a convective core introduces significant complexity into the modelling process. In particular, inaccuracies in the estimation of the surface correction coefficients ($a_3$ and $a_{-1}$) by the \citet{Ball&Gizon2014} prescription can propagate through the model by compensatory adjustments in other free parameters, such as initial helium and metal fractions ($Y_0$, $Z_0$), overshooting parameters, mass, and age. These adjustments, in turn, affect the structure and properties of the convective core, which then influences the global stellar structure. In essence, the presence of a convective core acts as a problem amplifier, magnifying the impact of even minor modelling inaccuracies and underscoring the need for surface-independent approaches in this regime.
 
In the lower panel of Fig.~\ref{fig:relative_differences_SE}, we present kernel density estimates of the relative differences in mass (blue), radius (orange), and age (green). Consistent with previous studies, our results confirm that the direct fitting of individual frequencies generally leads to an overestimation of stellar mass and radius, and an underestimation of stellar age, compared to surface-independent methods. On average, we found differences of 2.4\%, 0.8\%, and -8.0\%, for mass, radius, and age, respectively.

\subsection{Precision of stellar parameters}
\label{sec:precision_stellar_parameters}
For a subset of targets, the interpolation proved less effective, likely leading to underestimated uncertainties. Despite the high resolution of the stellar grid, which ensures the overall accuracy of the results, the MCMC sampling tends to cluster around grid points in those cases. This behaviour produces more sharply peaked posterior distributions, resulting in an overestimation of the precision. We identified this issue in eight targets, namely KIC3656476, KIC4914923, KIC6603624, KIC8694723, KIC8760414, KIC11853905, KIC6521045, and KIC12069449, which were consequently excluded from the precision analysis presented in this section.

Table~\ref{tab:precision_FICO} summarises the average statistical precision achieved for the LEGACY and Davies samples, as well as for the full catalogue. On (unweighted) average, the FICO procedure yields a statistical precision of 0.49\% in density, 2.3\% in mass, 0.82\% in radius, and 6.9\% in age, comfortably within the PLATO mission requirements\footnote{Formally speaking, PLATO requirements are accuracy requirements, not precision requirements. However, the statistical precision of the modelling procedure should remain (significantly) smaller than these requirements such that when combining statistical and bias uncertainties, the mission quality requirements are still satisfied.}. Fig.~\ref{fig:kde_precision} shows the kernel density estimates of the fractional uncertainty distribution for the stellar mass (blue), radius (orange), age (green), and mean density (red). For comparison, previous studies reported the following precisions: \citet{Silva-Aguirre2015} obtained 1.7\% (density), 3.3\% (mass), 1.2\% (radius), and 14\% (age) for the \citet{Davies2016} sample; \citet{Silva-Aguirre2017} reported 4\% (mass), 2\% (radius), and 10\% (age) for the LEGACY sample \citep{Lund2017}; and \citet{Bellinger2019d} achieved 3.5\% (mass), 1.4\% (radius), and 11\% (age) for a very similar catalogue to ours. These studies employed earlier-generation algorithms, and some of them have since undergone significant improvements \citep[e.g.][]{AguirreBorsen-Koch2022}. To our knowledge, updated analyses of the LEGACY and \citet{Davies2016} samples using the newer algorithms have not yet been published.

The FICO procedure incorporates the following improvements over previous methodologies:
\begin{enumerate}
\item We did not alter the uncertainties of the seismic constraints (no weights approach, see Sect. \ref{sec:FICO_modelling_procedure})
\item An improved interpolation scheme coupled with a higher-density grid of stellar models,
\item Inclusion of absolute luminosity as a classical constraint,
\item Use of the lowest radial-order frequency as an additional seismic constraint for targets with lower-quality convergence,
\item Incorporation of the inverted mean density as a classical constraint (which is a specificity of the FICO procedure),
\item A more intensive sampling strategy, allowing for extended burn-in periods and more robust convergence,
\item Treatment of the initial helium mass fraction as a free parameter (alongside initial metallicity), which can aid convergence in certain cases.
\end{enumerate}
It is important to note that each of the aforementioned studies \citep{Silva-Aguirre2015,Silva-Aguirre2017,Bellinger2019d} has its own methodological specificities, and not all improvements listed above apply simultaneously. We also note that it was difficult to find detailed information about the weighting used in some of the previous studies. Nevertheless, the combined effect of these enhancements contributes significantly to the improved precision observed in our results.

Based on the conclusions of \citet{Betrisey2023_AMS_surf}, which also apply to our sample, the inclusion of specific constraints appears to play a significant role in improving precision. In particular, the use of the inverted mean density and the radial frequency of the lowest radial order were identified as key contributors. It is important to note that the latter constraint is a double-edged sword: while it can enhance convergence for certain targets, \citet{Betrisey2023_AMS_surf} cautioned against its use in automated pipelines due to the potential for introducing substantial biases. In our case, we assessed the convergence of each minimisation individually, which allowed us to safely incorporate this constraint. As will be discussed in greater detail in the second paper of this series, the situation is more nuanced than the initial conclusions of \citet{Betrisey2023_AMS_surf}, which were based on a more limited dataset. 
The weighting approach also plays an important role as seismic constraints benefit from a high observational precision that is not artificially degraded in our study. The inclusion of the luminosity also contributes to the improved precision, albeit to a lesser extent. Other enhancements also contribute to the improved precision, though their individual impact appears more limited. However, their cumulative effect is likely more substantial. Quantifying this combined impact would require a dedicated analysis beyond the scope of the present study.

To further validate the consistency of our results, we examined empirical rules of thumb between the relative uncertainties in mass, radius, and age averaged over the sample. Pressure modes provide a tight constraint on the mean stellar density \citep[e.g.][]{Vandakurov1967}, implying that the relative mass uncertainty should, on average, be roughly three times larger than the relative radius uncertainty. The interpretation of age uncertainties along the main sequence is more complex. Ages are constrained primarily through the fractional evolutionary stage and the main-sequence lifetime, the latter being highly sensitive to both mass and composition. As a result, one may roughly approximate $\sigma_{\rm Age}/{\rm Age}$ to scale with a few times $\sigma_M/M$. Previous studies obtained empirical factors between 2 and 4 \citep{Silva-Aguirre2015,Silva-Aguirre2017,Bellinger2019d}. While such rules of thumb are valuable for sanity checks (averaged scaling factors of 1 or 10 would be alarming) they should not be applied in other contexts or on an individual star basis. In particular, the apparent age-mass trend must not be interpreted as a predictive scaling relation capable of estimating age uncertainties from mass uncertainties, especially for stars near the zero-age main sequence (ZAMS). It is only a rough approximation that holds when averaging relative uncertainties across an ensemble of stars distributed along the main sequence. We also note that our sample contains only one star close to the ZAMS (with $X_c > 0.6$), namely KIC 10644253. As expected, this target does not follow the global trend. The remaining stars, being more evolved along the MS, exhibit considerable scatter in the age-mass plane, as illustrated in Fig.~\ref{fig:slope_error}, yet an underlying trend remains discernible. We performed linear regressions on the relative uncertainties in the age-mass and mass-radius planes (see Appendix~\ref{sec:app:supplementary_data}), yielding:
\begin{align} 
\frac{\sigma_{\mathrm{Age}}}{\mathrm{Age}} &\sim (2.83 \pm 0.14)\frac{\sigma_M}{M}, \\
\frac{\sigma_M}{M} &\sim (2.84 \pm 0.02)\frac{\sigma_R}{R}, 
\end{align} 
which are consistent with empirical observations from previous studies.

\begin{table}[t!]
\caption{Average statistical precision of the FICO procedure.}
\resizebox{\linewidth}{!}{
\begin{tabular}{lcccc}
\hline \hline 
Sample & Mass & Radius & Age & Mean density \\ 
\hline 
\citet{Lund2017} & 2.2\% & 0.78\% & 6.6\% & 0.49\% \\
\citet{Davies2016} & 2.6\% & 0.92\% & 7.6\% & 0.52\% \\
Full catalogue & 2.3\% & 0.82\% & 6.9\% & 0.49\% \\
\hline
\end{tabular}
}
\label{tab:precision_FICO}
\end{table}

\begin{figure}[t!]
\centering
\includegraphics[scale=0.53]{Figures/kde_precision.png} 
\caption{Kernel density estimate of the fractional uncertainty distribution for the stellar mass (blue), radius (orange), age (green), and mean density (red).}
\label{fig:kde_precision}
\end{figure}

\subsection{Comparison with Silva-Aguirre et al. (2017)}
\label{sec:comparison_SA17}

\begin{figure*}[t!]
\centering
\begin{subfigure}[b]{.32\textwidth}
  \includegraphics[width=.99\textwidth]{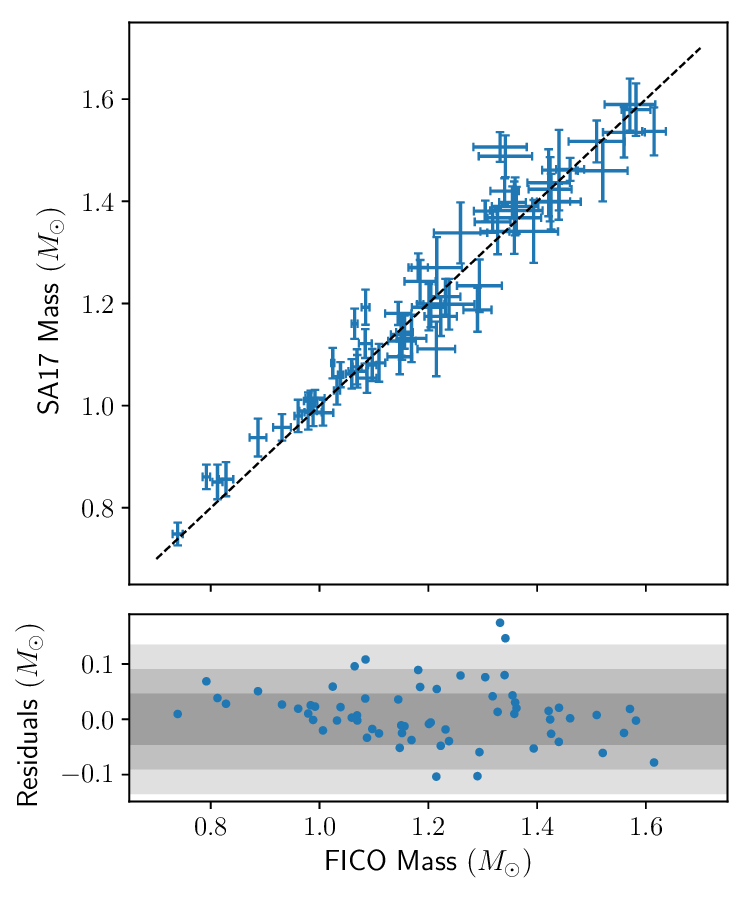}  
\end{subfigure}
\begin{subfigure}[b]{.32\textwidth}
  \includegraphics[width=.99\textwidth]{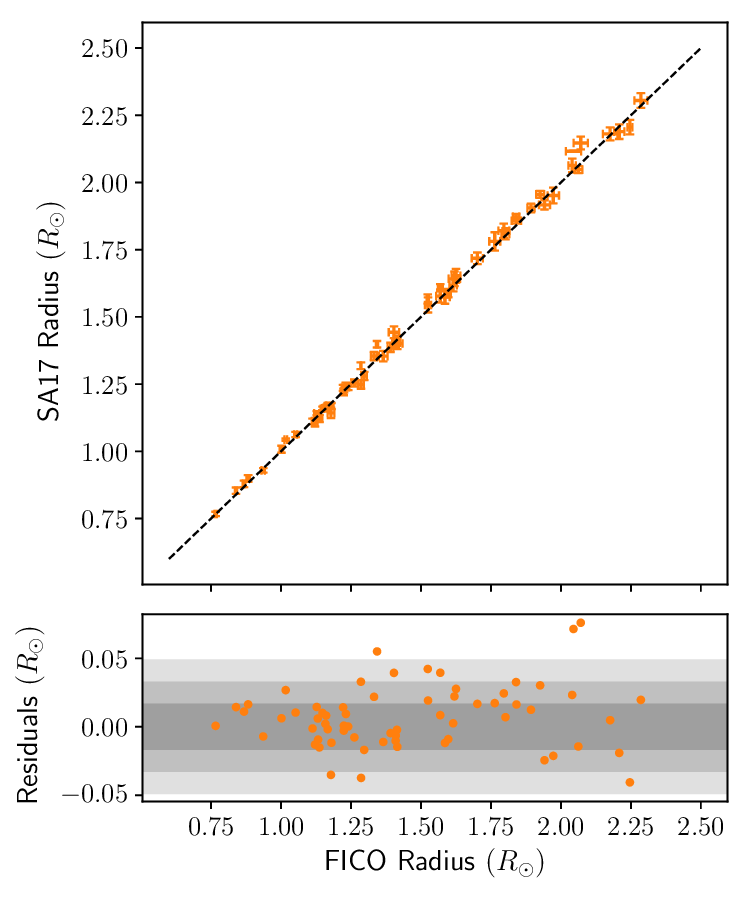}  
\end{subfigure}
\begin{subfigure}[b]{.32\textwidth}
  \includegraphics[width=.99\textwidth]{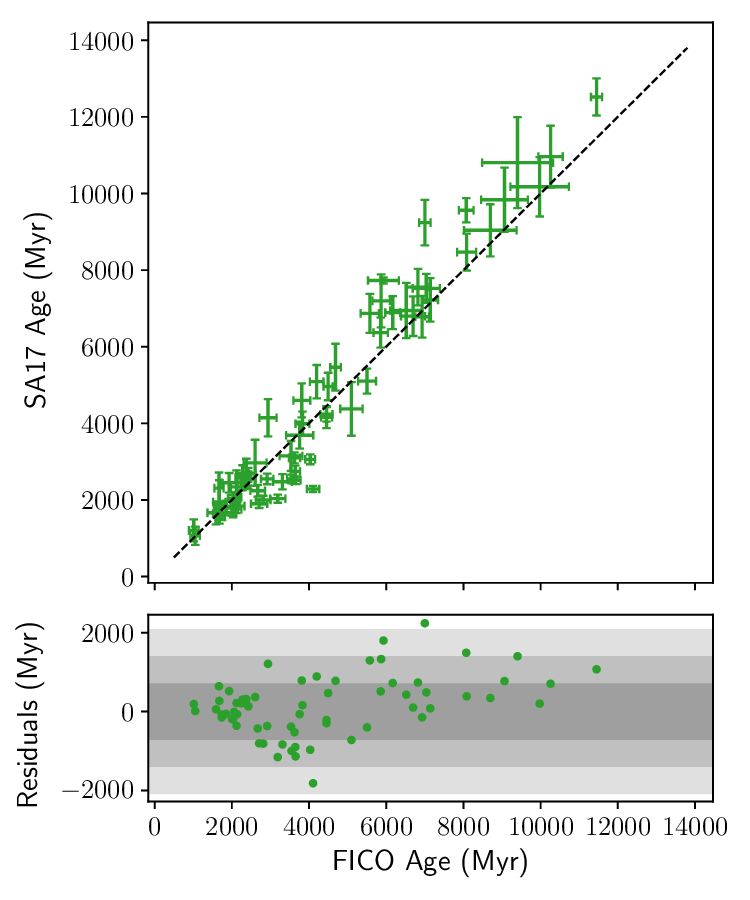}  
\end{subfigure}
\caption{Comparison between the FICO results and those of \citet{Silva-Aguirre2017} for the LEGACY sample. The upper panel presents a mirror plot contrasting the two sets of stellar parameters, while the lower panel illustrates the corresponding residuals. Shaded regions indicate the 1$\sigma$ (dark grey), 2$\sigma$ (medium grey), and 3$\sigma$ (light grey) intervals of the residual distributions. \textit{From left to right:} stellar mass, radius, and age.}
\label{fig:comparision_with_SA17}
\end{figure*}

In this section, we present a comparative analysis between our results for the LEGACY sample and those reported by \citet{Silva-Aguirre2017}. Fig.~\ref{fig:comparision_with_SA17} displays side-by-side plots of stellar mass, radius, and age. At first order, our findings are in good agreement with those of \citet{Silva-Aguirre2017}. The mean residuals, $(0.019 \pm 0.044)\, M_\odot$, $(0.008 \pm 0.016)\, R_\odot$, and $(-308 \pm 689)$ Myr, are statistically consistent with zero, indicating no significant systematic offset.

However, a second-order inspection, using relative differences, reveals subtle trends: \citet{Silva-Aguirre2017} tend to infer slightly larger masses and radii, which in turn yield younger stellar ages. On average, we observed relative biases of 1.5\%, 0.7\% and -11.5\%, respectively. We primarily attribute these systematic discrepancies to differences in the adopted physical inputs, the choice of the optimised variables (notably, in our study, the initial helium mass fraction is free as well as the overshooting parameter for targets with $M>1.1M_\odot$), and the performance of the optimisation strategies. While both studies use the same seismic datasets in principle, we note that for certain targets, we excluded anomalous modes that visibly deviate in the observational échelle diagram, also potentially contributing to the observed divergences.

Our models assumed a low-metallicity tables \citep{Asplund2009}, whereas \citet{Silva-Aguirre2017} employed high-metallicity values \citep{Grevesse&Sauval1998}. Metallicity can exert a significant influence on stellar structure and evolution, primarily through its impact on opacity. An increase in metal content enhances the opacity of stellar material, thereby diminishing the efficiency of radiative energy transport \citep[see e.g.][for a textbook]{Salaris&Cassisi2005}. For a given mass, this results in cooler, more extended stellar envelopes, leading to larger radii and lower effective temperatures compared to their low-metallicity counterparts. When fitting seismic and spectroscopic observables, these structural disparities propagate into the inferred global parameters. High-metallicity models typically require larger masses and radii to reproduce the same observed frequencies and luminosities. This mass adjustment has a pronounced effect on the inferred age. Although higher metallicity alone tends to slow down stellar evolution, owing to reduced core temperatures and slower nuclear burning rates, the accompanying increase in mass accelerates evolutionary timescales significantly. The latter effect dominates, leading high-metallicity solutions to yield younger ages for identical observational constraints.

Naturally, the performance of the optimisation algorithm contributes to the scatter in the results, as reflected in the large standard deviation of the residuals. The results presented by \citet{Silva-Aguirre2017} are based on earlier versions of GARSTEC and BASTA, whereas our methodology leverages the latest iteration of AIMS, integrated with a high-resolution grid of stellar models and additional constraints, as detailed in the preceding section. Despite these methodological differences, the average biases align well with theoretical expectations, thereby reinforcing the reliability and robustness of our modelling framework.

Furthermore, we note that the average relative bias in stellar mass is smaller than the statistical precision achieved by the FICO procedure (1.5\% versus 2.3\%), while the radius bias is comparable (0.7\% versus 0.8\%). In contrast, the age bias exceeds the statistical precision (11.5\% versus 6.9\%). Importantly, both the mass and radius biases fall comfortably within the PLATO mission requirements. However, the age bias is comparable to the 10\% threshold of PLATO, underscoring the significant influence that the choice of physical ingredients in stellar models can exert, particularly at higher stellar masses.

To identify cases of incompatibility between our results and those of \citet{Silva-Aguirre2017}, we defined normalised residuals, denoted as $\beta$, as the absolute value of the standard residual divided by the sum of the statistical uncertainties. This parameter quantifies the overlap between the uncertainty intervals of two variables. For example, a value of $\beta < 1$ indicates that the $1\sigma$-intervals overlap. A value of $\beta > 3$ was assumed to indicate incompatibility in the stellar characterisations. Table~\ref{tab:classification_residuals} summarises the compatibility levels based on this classification. We identified ten incompatible cases: two based on radius comparisons and eight on age.

Among these, two stars (KIC7106245 and KIC8760414) in the \citet{Silva-Aguirre2017} dataset exhibit initial helium abundances below 0.21, which we interpret as indicative of convergence towards non-physical solutions. KIC9353712 suffers from low-quality seismic data, with small oscillation peaks compared to the noise level, and it lies in a higher-degeneracy region of parameter space \citep[see e.g.][]{White2011}, which likely explains the discrepancies. For KIC3632418, KIC3656476, KIC7940546, KIC8228742, KIC10068307, and KIC10162436, the peak-bagging plots in \citet{Lund2017} seem to indicate that their algorithm had more difficulties in distinguishing quadrupolar from radial modes, which lowers the reliability of the extracted $l=2$ modes, and possible of the $l=0$ modes as well. These stars also reside in regions of elevated degeneracy. KIC4914923 is more puzzling. The FICO procedure exhibited stable convergence, and there is no evident indication of convergence issues in the \citet{Silva-Aguirre2017} analysis. It is plausible that variations in chemical abundances exert a more pronounced influence on this particular star. It could also be the sign of the presence of non-standard physical processes (e.g. turbulent diffusion, undershooting, radiative accelerations) not accounted for in our models and those of \citet{Silva-Aguirre2017}. However, a detailed, target-specific investigation would be required to draw firmer conclusions.

\begin{table}[t!]
\centering
\caption{Compatibility classification of the FICO and \citet{Silva-Aguirre2015,Silva-Aguirre2017} stellar parameters.}
\resizebox{\linewidth}{!}{
\begin{tabular}{lccc}
\hline \hline 
Classification & Mass & Radius & Age \\ 
\hline 
\multicolumn{4}{l}{\textit{\citet{Lund2017}}} \\
$\beta \leq 1$ & $53/65 = 81.5\%$ & $ 52/65 = 80.0\%$ & $39/65 = 60.0\%$ \\
$\beta \leq 2$ & $61/65 = 93.8\%$ & $60/65 = 92.3\%$ & $54/65 = 83.1\%$ \\
$\beta \leq 3$ & $65/65 = 100\%$ & $63/65 = 96.9\%$ & $57/65 = 87.7\%$ \\
$\beta > 3$ & $0/65 = 0.0\%$ & $2/65 = 3.1\%$ & $8/65 = 12.3\%$ \\
\hline
\multicolumn{4}{l}{\textit{\citet{Davies2016}}} \\
$\beta \leq 1$ & $18/29 = 62.1\%$ & $18/29 = 62.1\%$ & $18/29 = 62.1\%$ \\
$\beta \leq 2$ & $25/29 = 86.2\%$ & $26/29 = 89.7\%$ & $23/29 = 79.3\%$ \\
$\beta \leq 3$ & $27/29 = 93.1\%$ & $27/29 = 93.1\%$ & $27/29 = 93.1\%$ \\
$\beta > 3$ & $2/29 = 6.9\%$ & $2/29 = 6.9\%$ & $2/29 = 6.9\%$ \\
\hline 
\end{tabular}
}
\label{tab:classification_residuals} 
\end{table}

\subsection{Comparison with Silva-Aguirre et al. (2015)}
\label{sec:comparison_SA15}

\begin{figure}[t!]
\centering
\includegraphics[scale=0.5]{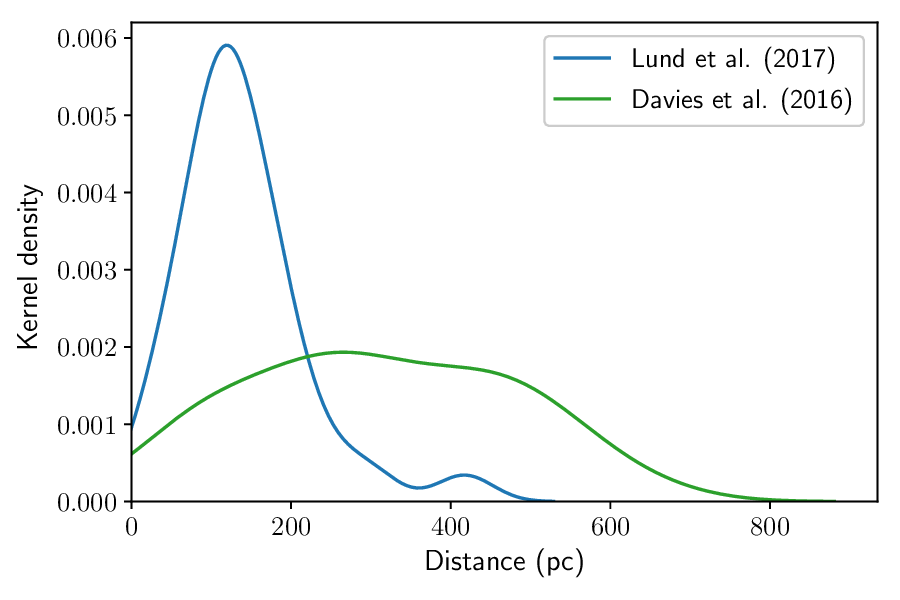} 
\caption{Kernel density estimate of the distance distribution for the LEGACY (blue) and \citet{Davies2016} (green) samples.}
\label{fig:kde_distances}
\end{figure}

In this section, we undertake a comparative assessment of our results for the \citet{Davies2016} sample against those reported by \citet{Silva-Aguirre2015}. Unlike the LEGACY sample discussed previously, this comparison presents greater challenges for quantitative analysis, owing to two principal factors. Firstly, the median distance of the \citet{Davies2016} targets is significantly greater than that of the LEGACY sample, 299 pc compared to 123 pc, as illustrated in Fig.~\ref{fig:kde_distances}. This increased distance results in a lower signal-to-noise ratio, rendering the extracted oscillation frequencies less robust. In essence, while the LEGACY sample benefits from high-quality data, the Davies targets are characterised by moderate data quality. Secondly, the statistical strength of the sample is limited. This, combined with the fact that the fundamental stellar parameters in \citet{Silva-Aguirre2015} were derived using an early version of the BASTA software, contributes to the considerable scatter observed in Fig.~\ref{fig:comparision_with_SA15} and Table~\ref{tab:classification_residuals}.

Nonetheless, we can confidently affirm that the average residuals for stellar mass, radius, and age are all consistent with zero, mostly thanks to the large standard deviation. Notably, the majority of discrepant cases between our findings and those of \citet{Silva-Aguirre2015} are concentrated in the mass range between 1.2 and 1.4 solar masses. Furthermore, based on the normalised residuals $\beta$, we identified three targets as incompatible: KIC8554498, KIC10514430, and KIC11853905. Additional visual inspection of Fig.~\ref{fig:comparision_with_SA15} also revealed that KIC4349452 is an outlier. KIC10514430 exhibits low data quality, which likely accounts for the observed divergence. In the cases of KIC4349452, KIC8554498 and KIC11853905, the noise levels significantly exceed the amplitudes of the fitted oscillation frequencies, making it difficult to visually discern the acoustic power excess in the power spectrum. Consequently, the reliability of the detected modes is diminished, and our minimisation algorithm converged on solutions that differ from those reported by \citet{Silva-Aguirre2015}.

\section{Conclusions}
\label{sec:conclusions}
In this study, we presented detailed asteroseismic modelling of 95 main-sequence solar-like stars and \textit{Kepler} exoplanet host stars using the FICO procedure. We began in Sect.~\ref{sec:presentation_stellar_catalogue} by introducing the stellar catalogue that formed the foundation of our analysis, along with the seismic and non-seismic observational constraints that were considered. Sect.~\ref{sec:modelling_strategy} described the modelling strategy adopted to perform the asteroseismic characterisation and to derive key stellar parameters, including mass, radius, age, and mean density. Sect.~\ref{sec:results} presented the results of our asteroseismic analysis, starting with the Sun (using degraded data) as a control target, followed by the full catalogue. A comprehensive list of the derived stellar parameters is provided in Appendix~\ref{sec:app:complete_fundamental_stellar_data}. In Sect.~\ref{sec:discussion}, we presented a detailed discussion of our findings, focusing on three main aspects: (i) a comparison between our results, obtained via the FICO procedure which relies on surface-independent constraints, and a direct fit to individual frequencies using a semi-empirical treatment of surface effects; (ii) an in-depth evaluation of the precision performances of our methodology; and (iii) a comparison with values reported in the literature.

The FICO procedure \citep{Betrisey2022,Betrisey2023_AMS_surf,Betrisey2024_AMS_quality,Betrisey2024_phd} demonstrated excellent performance across a broad sample comprising just under one hundred main-sequence solar-like stars. Complementary consistency tests to \citet{Betrisey2023_AMS_surf}, who investigated the accuracy of FICO based on synthetic targets and well-studied calibrators, using degraded solar data re-confirmed the robustness of our methodology, with derived parameters closely matching expected solar values. As an aside, we note that the number of benchmark stars with independent mass and radius observations currently remains limited, a situation that PLATO is expected to improve thanks to the likely overlap between its long‑duration observing field(s) and targets with interferometric or binary‑based measurements. Furthermore, the FICO procedure achieved, using a no weights approach for the classical and seismic constraints (notably, the uncertainty of the oscillation frequencies is not altered to overweight classical constraints), average statistical precisions of 2.3\%, 0.82\%, 6.9\%, and 0.49\% for stellar mass, radius, age, and mean density, respectively. The precision performance of the FICO procedure is therefore compatible with the requirements set by the PLATO mission \citep[15\% in mass, 1-2\% in radius, and 10\% in age for Sun-like stars; see][]{Rauer2025}.

From a statistical standpoint, our results were generally in good agreement with stellar parameters inferred from the literature \citep{Silva-Aguirre2015,Silva-Aguirre2017}. However, at PLATO-level precision, we detected non-negligible systematic differences, particularly in stellar age. We believe that these discrepancies may primarily be attributed to the performance of the minimisation algorithms, the results of \citet{Silva-Aguirre2015,Silva-Aguirre2017} being based on old versions of GARSTEC and BASTA, the impact of the number of free parameters (e.g. overshooting and initial helium mass fraction), a possible presence of non-standard physical processes (e.g. turbulent diffusion, undershooting, radiative accelerations) not accounted for in our models and those of \citet{Silva-Aguirre2015,Silva-Aguirre2017}, especially for the more massive targets, and the choice of initial chemical abundances: we adopted the low-metallicity tables of \citet{Asplund2009}, whereas the literature values were based on the high-metallicity \citet{Grevesse&Sauval1998} tables. These findings underscored the importance of carefully selecting and benchmarking model physics, especially in the context of the PLATO mission, where high-precision stellar characterisation is essential for accurate characterisation of exoplanetary systems. Moreover, chemical abundances are not the only physical ingredient that can significantly influence asteroseismic characterisation \citep[e.g.][]{Buldgen2019f,Farnir2020,Betrisey2022}. For example, \citet{Betrisey2022} highlighted the impact of opacities on the characterisation of the Sun-like star Kepler-93. This highlights a broader challenge. The physics used in asteroseismic models is deeply rooted in solar physics, and the solar modelling problem \citep[see e.g.][]{Basu&Antia2008,Buldgen2019e}, referring to the persistent mismatch between theoretical solar models and helioseismic observations, remains unresolved. Recent opacity measurements \citep{Bailey2015,Nagayama2019,Buldgen2025_opacities} have shown that a revision of radiative opacities in stellar conditions would improve the situation. However, it is also likely that additional ingredients (e.g. transport of chemical, equation of state) of stellar evolution models might be revised to restore the agreement \citep{Buldgen2024_opacities,Buldgen2025_wishlist}. Such revisions would also likely impact the results of asteroseismic modelling.

Our analysis of the treatment of surface effects further strengthened the conclusions drawn by \citet{Betrisey2023_AMS_surf}, who advocated for the systematic use of surface-independent methods above a certain stellar mass threshold. Although direct frequency fitting with semi-empirical corrections remains a common practice, our results clearly demonstrate that this approach tends to overestimate stellar mass and radius, while underestimating age, consistently with literature findings on smaller samples \citep[e.g.][]{Ball&Gizon2017,Nsamba2018,Jorgensen2020,Jorgensen2021,Cunha2021,Betrisey2023_AMS_surf}. In contrast, the FICO procedure, by leveraging inversion techniques and frequency separation ratios, effectively mitigates these biases, yielding more reliable stellar characterisations. At PLATO-level precision, we identified two distinct regimes. In near-solar conditions, both modelling approaches performed comparably. However, this equivalence breaks down if the stellar properties depart from solar conditions. We observed a regime transition around $1.20M_\odot$ or 6200~K, beyond which surface-independent methods consistently outperformed traditional approaches. Given the gradual nature of this transition, we recommend systematically adopting surface-independent techniques for stars with masses exceeding $1.15M_\odot$ or with effective temperatures above 6050~K, ensuring unbiased parameter estimation.

As the launch of PLATO approaches, the development of robust and scalable modelling techniques by several independent teams becomes increasingly important. PLATO is expected to expand the pool of high-quality seismic observations of main-sequence solar-like stars from approximately one hundred to between 800 and 1000 \citep{Goupil2024}, marking a significant shift in paradigm. These enhanced statistics will enable comprehensive statistical and population studies that were previously unattainable. In this evolving context, the FICO procedure, characterised by its hybrid forward-inverse architecture, stands out as a promising approach for the precise and accurate characterisation of exoplanet host stars in the PLATO era.

\begin{acknowledgements}
We thank Dr. Gaël Buldgen and Dr. Saniya Khan for fruitful discussions and constructive suggestions. J.B. acknowledges funding from the SNF Postdoc.Mobility grant no. P500PT{\_}222217 (Impact of magnetic activity on the characterization of FGKM main-sequence host-stars). C.P. thanks the Belgian Federal Science Policy Office (BEL-SPO) for their financial support in the framework of the PRODEX Program of the European Space Agency (ESA) under contract number 4000141194. M.S.C. acknowledges support by Fundação para a Ciência e a Tecnologia (FCT) through the research grant UID/04434/2025 and the Investigador FCT Contract with reference 2023.09303.CEECIND/CP2839/CT0003. Finally, this work has benefited from financial support by CNES (Centre National des Études Spatiales) in the framework of its contribution to the PLATO mission.
\end{acknowledgements}

\bibliography{bibliography.bib}

\begin{thebibliography}{128}
\expandafter\ifx\csname natexlab\endcsname\relax\def\natexlab#1{#1}\fi

\bibitem[{{Aerts}(2021)}]{Aerts2021}
{Aerts}, C. 2021, Reviews of Modern Physics, 93, 015001

\bibitem[{{Aguirre B{\o}rsen-Koch} {et~al.}(2022){Aguirre B{\o}rsen-Koch},
  {R{\o}rsted}, {Justesen}, {Stokholm}, {Verma}, {Winther}, {Knudstrup},
  {Nielsen}, {Sahlholdt}, {Larsen}, {Cassisi}, {Serenelli}, {Casagrande},
  {Christensen-Dalsgaard}, {Davies}, {Ferguson}, {Lund}, {Weiss}, \&
  {White}}]{AguirreBorsen-Koch2022}
{Aguirre B{\o}rsen-Koch}, V., {R{\o}rsted}, J.~L., {Justesen}, A.~B., {et~al.}
  2022, \mnras, 509, 4344

\bibitem[{{Ahuir} {et~al.}(2021){Ahuir}, {Mathis}, \& {Amard}}]{Ahuir2021}
{Ahuir}, J., {Mathis}, S., \& {Amard}, L. 2021, \aap, 651, A3

\bibitem[{{Appourchaux}(2014)}]{Appourchaux2014}
{Appourchaux}, T. 2014, in Asteroseismology, ed. P.~L. {Pall{\'e}} \&
  C.~{Esteban}, 123

\bibitem[{{Appourchaux} {et~al.}(2012){Appourchaux}, {Chaplin}, {Garc{\'\i}a},
  {Gruberbauer}, {Verner}, {Antia}, {Benomar}, {Campante}, {Davies},
  {Deheuvels}, {Handberg}, {Hekker}, {Howe}, {R{\'e}gulo}, {Salabert},
  {Bedding}, {White}, {Ballot}, {Mathur}, {Silva Aguirre}, {Elsworth}, {Basu},
  {Gilliland}, {Christensen-Dalsgaard}, {Kjeldsen}, {Uddin}, {Stumpe}, \&
  {Barclay}}]{Appourchaux2012}
{Appourchaux}, T., {Chaplin}, W.~J., {Garc{\'\i}a}, R.~A., {et~al.} 2012, \aap,
  543, A54

\bibitem[{{Appourchaux} {et~al.}(1998){Appourchaux}, {Gizon}, \&
  {Rabello-Soares}}]{Appourchaux1998}
{Appourchaux}, T., {Gizon}, L., \& {Rabello-Soares}, M.-C. 1998, \aaps, 132,
  107

\bibitem[{{Asplund} {et~al.}(2021){Asplund}, {Amarsi}, \&
  {Grevesse}}]{Asplund2021}
{Asplund}, M., {Amarsi}, A.~M., \& {Grevesse}, N. 2021, \aap, 653, A141

\bibitem[{{Asplund} {et~al.}(2009){Asplund}, {Grevesse}, {Sauval}, \&
  {Scott}}]{Asplund2009}
{Asplund}, M., {Grevesse}, N., {Sauval}, A.~J., \& {Scott}, P. 2009, \araa, 47,
  481

\bibitem[{{Baglin} {et~al.}(2009){Baglin}, {Auvergne}, {Barge}, {Deleuil},
  {Michel}, \& {CoRoT Exoplanet Science Team}}]{Baglin2009}
{Baglin}, A., {Auvergne}, M., {Barge}, P., {et~al.} 2009, in IAU Symposium,
  Vol. 253, Transiting Planets, ed. F.~{Pont}, D.~{Sasselov}, \& M.~J.
  {Holman}, 71--81

\bibitem[{{Bailer-Jones} {et~al.}(2021){Bailer-Jones}, {Rybizki}, {Fouesneau},
  {Demleitner}, \& {Andrae}}]{Bailer-Jones2021}
{Bailer-Jones}, C.~A.~L., {Rybizki}, J., {Fouesneau}, M., {Demleitner}, M., \&
  {Andrae}, R. 2021, \aj, 161, 147

\bibitem[{{Bailey} {et~al.}(2015){Bailey}, {Nagayama}, {Loisel}, {Rochau},
  {Blancard}, {Colgan}, {Cosse}, {Faussurier}, {Fontes}, {Gilleron},
  {Golovkin}, {Hansen}, {Iglesias}, {Kilcrease}, {Macfarlane}, {Mancini},
  {Nahar}, {Orban}, {Pain}, {Pradhan}, {Sherrill}, \& {Wilson}}]{Bailey2015}
{Bailey}, J.~E., {Nagayama}, T., {Loisel}, G.~P., {et~al.} 2015, \nat, 517, 56

\bibitem[{{Ball} \& {Gizon}(2014)}]{Ball&Gizon2014}
{Ball}, W.~H. \& {Gizon}, L. 2014, \aap, 568, A123

\bibitem[{{Ball} \& {Gizon}(2017)}]{Ball&Gizon2017}
{Ball}, W.~H. \& {Gizon}, L. 2017, \aap, 600, A128

\bibitem[{{Barker}(2020)}]{Barker2020}
{Barker}, A.~J. 2020, \mnras, 498, 2270

\bibitem[{{Basu} \& {Antia}(2008)}]{Basu&Antia2008}
{Basu}, S. \& {Antia}, H.~M. 2008, \physrep, 457, 217

\bibitem[{{Bazot} {et~al.}(2008){Bazot}, {Bourguignon}, \&
  {Christensen-Dalsgaard}}]{Bazot2008}
{Bazot}, M., {Bourguignon}, S., \& {Christensen-Dalsgaard}, J. 2008, \memsai,
  79, 660

\bibitem[{{Bellinger} {et~al.}(2016){Bellinger}, {Angelou}, {Hekker}, {Basu},
  {Ball}, \& {Guggenberger}}]{Bellinger2016}
{Bellinger}, E.~P., {Angelou}, G.~C., {Hekker}, S., {et~al.} 2016, \apj, 830,
  31

\bibitem[{{Bellinger} {et~al.}(2019){Bellinger}, {Hekker}, {Angelou},
  {Stokholm}, \& {Basu}}]{Bellinger2019d}
{Bellinger}, E.~P., {Hekker}, S., {Angelou}, G.~C., {Stokholm}, A., \& {Basu},
  S. 2019, \aap, 622, A130

\bibitem[{{B{\'e}trisey}(2024)}]{Betrisey2024_phd}
{B{\'e}trisey}, J. 2024, PhD thesis, University of Geneva, Switzerland

\bibitem[{{B{\'e}trisey} {et~al.}(2025{\natexlab{a}}){B{\'e}trisey},
  {Broomhall}, {Breton}, {Garc{\'\i}a}, {Davenport}, \&
  {Kochukhov}}]{Betrisey2025_MA_Damping}
{B{\'e}trisey}, J., {Broomhall}, A.~M., {Breton}, S.~N., {et~al.}
  2025{\natexlab{a}}, \aap, 699, L9

\bibitem[{{B{\'e}trisey} \& {Buldgen}(2022)}]{Betrisey&Buldgen2022}
{B{\'e}trisey}, J. \& {Buldgen}, G. 2022, \aap, 663, A92

\bibitem[{{B{\'e}trisey} {et~al.}(2023{\natexlab{a}}){B{\'e}trisey}, {Buldgen},
  {Reese}, {Farnir}, {Dupret}, {Khan}, {Goupil}, {Eggenberger}, \&
  {Meynet}}]{Betrisey2023_AMS_surf}
{B{\'e}trisey}, J., {Buldgen}, G., {Reese}, D.~R., {et~al.} 2023{\natexlab{a}},
  \aap, 676, A10

\bibitem[{{B{\'e}trisey} {et~al.}(2024{\natexlab{a}}){B{\'e}trisey}, {Buldgen},
  {Reese}, \& {Meynet}}]{Betrisey2024_AMS_quality}
{B{\'e}trisey}, J., {Buldgen}, G., {Reese}, D.~R., \& {Meynet}, G.
  2024{\natexlab{a}}, \aap, 681, A99

\bibitem[{{B{\'e}trisey} {et~al.}(2023{\natexlab{b}}){B{\'e}trisey},
  {Eggenberger}, {Buldgen}, {Benomar}, \& {Bazot}}]{Betrisey2023_rot}
{B{\'e}trisey}, J., {Eggenberger}, P., {Buldgen}, G., {Benomar}, O., \&
  {Bazot}, M. 2023{\natexlab{b}}, \aap, 673, L11

\bibitem[{{B{\'e}trisey} {et~al.}(2024{\natexlab{b}}){B{\'e}trisey}, {Farnir},
  {Breton}, {Garc{\'\i}a}, {Broomhall}, {Amarsi}, \&
  {Kochukhov}}]{Betrisey2024_MA_Sun}
{B{\'e}trisey}, J., {Farnir}, M., {Breton}, S.~N., {et~al.} 2024{\natexlab{b}},
  \aap, 688, L17

\bibitem[{{B{\'e}trisey} {et~al.}(2022){B{\'e}trisey}, {Pezzotti}, {Buldgen},
  {Khan}, {Eggenberger}, {Salmon}, \& {Miglio}}]{Betrisey2022}
{B{\'e}trisey}, J., {Pezzotti}, C., {Buldgen}, G., {et~al.} 2022, \aap, 659,
  A56

\bibitem[{{B{\'e}trisey} {et~al.}(2025{\natexlab{b}}){B{\'e}trisey}, {Reese},
  {Breton}, {Broomhall}, {Amarsi}, {Garc{\'\i}a}, \&
  {Kochukhov}}]{Betrisey2025_MA_Inv}
{B{\'e}trisey}, J., {Reese}, D.~R., {Breton}, S.~N., {et~al.}
  2025{\natexlab{b}}, \aap, 697, A219

\bibitem[{{Borucki} {et~al.}(2010){Borucki}, {Koch}, {Basri}, {Batalha},
  {Brown}, {Caldwell}, {Caldwell}, {Christensen-Dalsgaard}, {Cochran},
  {DeVore}, {Dunham}, {Dupree}, {Gautier}, {Geary}, {Gilliland}, {Gould},
  {Howell}, {Jenkins}, {Kondo}, {Latham}, {Marcy}, {Meibom}, {Kjeldsen},
  {Lissauer}, {Monet}, {Morrison}, {Sasselov}, {Tarter}, {Boss}, {Brownlee},
  {Owen}, {Buzasi}, {Charbonneau}, {Doyle}, {Fortney}, {Ford}, {Holman},
  {Seager}, {Steffen}, {Welsh}, {Rowe}, {Anderson}, {Buchhave}, {Ciardi},
  {Walkowicz}, {Sherry}, {Horch}, {Isaacson}, {Everett}, {Fischer}, {Torres},
  {Johnson}, {Endl}, {MacQueen}, {Bryson}, {Dotson}, {Haas}, {Kolodziejczak},
  {Van Cleve}, {Chandrasekaran}, {Twicken}, {Quintana}, {Clarke}, {Allen},
  {Li}, {Wu}, {Tenenbaum}, {Verner}, {Bruhweiler}, {Barnes}, \&
  {Prsa}}]{Borucki2010}
{Borucki}, W.~J., {Koch}, D., {Basri}, G., {et~al.} 2010, Science, 327, 977

\bibitem[{{Bowman} {et~al.}(2022){Bowman}, {Vandenbussche}, {Sana},
  {Tkachenko}, {Raskin}, {Delabie}, {Vandoren}, {Royer}, {Garcia}, {Van Reeth},
  \& {CubeSpec Collaboration}}]{Bowman2022}
{Bowman}, D.~M., {Vandenbussche}, B., {Sana}, H., {et~al.} 2022, \aap, 658, A96

\bibitem[{{Buldgen} {et~al.}(2025{\natexlab{a}}){Buldgen}, {B{\'e}trisey},
  {Pezzotti}, {Borisov}, \& {Noels}}]{Buldgen2025_lithium}
{Buldgen}, G., {B{\'e}trisey}, J., {Pezzotti}, C., {Borisov}, S., \& {Noels},
  A. 2025{\natexlab{a}}, \aap, 702, A162

\bibitem[{{Buldgen} {et~al.}(2022{\natexlab{a}}){Buldgen}, {B{\'e}trisey},
  {Roxburgh}, {Vorontsov}, \& {Reese}}]{Buldgen2022c}
{Buldgen}, G., {B{\'e}trisey}, J., {Roxburgh}, I.~W., {Vorontsov}, S.~V., \&
  {Reese}, D.~R. 2022{\natexlab{a}}, Frontiers in Astronomy and Space Sciences,
  9, 942373

\bibitem[{{Buldgen} {et~al.}(2025{\natexlab{b}}){Buldgen}, {Canocchi}, {Le
  Saux}, {Baturin}, {Trampedach}, {Oreshina}, {Ayukov}, {Pradhan}, {Pain},
  {Kunitomo}, {Appourchaux}, {Garc{\'\i}a}, {Deal}, {Grevesse}, {Noels},
  {Christensen-Dalsgaard}, {Guillot}, {Nandal}, {B{\'e}trisey}, {Blancard},
  {Colgan}, {Coss{\'e}}, {Fontes}, {Petitdemange}, \&
  {Pin{\c{c}}on}}]{Buldgen2025_wishlist}
{Buldgen}, G., {Canocchi}, G., {Le Saux}, A., {et~al.} 2025{\natexlab{b}},
  \solphys, 300, 97

\bibitem[{{Buldgen} {et~al.}(2022{\natexlab{b}}){Buldgen}, {Farnir},
  {Eggenberger}, {B{\'e}trisey}, {Pezzotti}, {Pin{\c{c}}on}, {Deal}, \&
  {Salmon}}]{Buldgen2022b}
{Buldgen}, G., {Farnir}, M., {Eggenberger}, P., {et~al.} 2022{\natexlab{b}},
  \aap, 661, A143

\bibitem[{{Buldgen} {et~al.}(2019{\natexlab{a}}){Buldgen}, {Farnir},
  {Pezzotti}, {Eggenberger}, {Salmon}, {Montalban}, {Ferguson}, {Khan},
  {Bourrier}, {Rendle}, {Meynet}, {Miglio}, \& {Noels}}]{Buldgen2019f}
{Buldgen}, G., {Farnir}, M., {Pezzotti}, C., {et~al.} 2019{\natexlab{a}}, \aap,
  630, A126

\bibitem[{{Buldgen} {et~al.}(2024){Buldgen}, {Noels}, {Scuflaire}, {Amarsi},
  {Grevesse}, {Eggenberger}, {Colgan}, {Fontes}, {Baturin}, {Oreshina},
  {Ayukov}, {Hakel}, \& {Kilcrease}}]{Buldgen2024_opacities}
{Buldgen}, G., {Noels}, A., {Scuflaire}, R., {et~al.} 2024, \aap, 686, A108

\bibitem[{{Buldgen} {et~al.}(2025{\natexlab{c}}){Buldgen}, {Pain}, {Coss{\'e}},
  {Blancard}, {Gilleron}, {Pradhan}, {Fontes}, {Colgan}, {Noels},
  {Christensen-Dalsgaard}, {Deal}, {Ayukov}, {Baturin}, {Oreshina},
  {Scuflaire}, {Pin{\c{c}}on}, {Lebreton}, {Corbard}, {Eggenberger}, {Hakel},
  \& {Kilcrease}}]{Buldgen2025_opacities}
{Buldgen}, G., {Pain}, J.-C., {Coss{\'e}}, P., {et~al.} 2025{\natexlab{c}},
  Nature Communications, 16, 693

\bibitem[{{Buldgen} {et~al.}(2017){Buldgen}, {Reese}, \&
  {Dupret}}]{Buldgen2017c}
{Buldgen}, G., {Reese}, D., \& {Dupret}, M.-A. 2017, in European Physical
  Journal Web of Conferences, Vol. 160, European Physical Journal Web of
  Conferences, 03005

\bibitem[{{Buldgen} {et~al.}(2015{\natexlab{a}}){Buldgen}, {Reese}, \&
  {Dupret}}]{Buldgen2015b}
{Buldgen}, G., {Reese}, D.~R., \& {Dupret}, M.~A. 2015{\natexlab{a}}, \aap,
  583, A62

\bibitem[{{Buldgen} {et~al.}(2016{\natexlab{a}}){Buldgen}, {Reese}, \&
  {Dupret}}]{Buldgen2016b}
{Buldgen}, G., {Reese}, D.~R., \& {Dupret}, M.~A. 2016{\natexlab{a}}, \aap,
  585, A109

\bibitem[{{Buldgen} {et~al.}(2015{\natexlab{b}}){Buldgen}, {Reese}, {Dupret},
  \& {Samadi}}]{Buldgen2015a}
{Buldgen}, G., {Reese}, D.~R., {Dupret}, M.~A., \& {Samadi}, R.
  2015{\natexlab{b}}, \aap, 574, A42

\bibitem[{{Buldgen} {et~al.}(2019{\natexlab{b}}){Buldgen}, {Rendle}, {Sonoi},
  {Davies}, {Miglio}, {Salmon}, {Reese}, {Bossini}, {Eggenberger}, {Noels}, \&
  {Scuflaire}}]{Buldgen2019b}
{Buldgen}, G., {Rendle}, B., {Sonoi}, T., {et~al.} 2019{\natexlab{b}}, \mnras,
  482, 2305

\bibitem[{{Buldgen} {et~al.}(2019{\natexlab{c}}){Buldgen}, {Salmon}, \&
  {Noels}}]{Buldgen2019e}
{Buldgen}, G., {Salmon}, S., \& {Noels}, A. 2019{\natexlab{c}}, Frontiers in
  Astronomy and Space Sciences, 6, 42

\bibitem[{{Buldgen} {et~al.}(2016{\natexlab{b}}){Buldgen}, {Salmon}, {Reese},
  \& {Dupret}}]{Buldgen2016c}
{Buldgen}, G., {Salmon}, S.~J.~A.~J., {Reese}, D.~R., \& {Dupret}, M.~A.
  2016{\natexlab{b}}, \aap, 596, A73

\bibitem[{{Casagrande} \& {VandenBerg}(2014)}]{Casagrande2014}
{Casagrande}, L. \& {VandenBerg}, D.~A. 2014, \mnras, 444, 392

\bibitem[{{Casagrande} \& {VandenBerg}(2018)}]{Casagrande2018}
{Casagrande}, L. \& {VandenBerg}, D.~A. 2018, \mnras, 475, 5023

\bibitem[{{Chaplin} \& {Miglio}(2013)}]{Chaplin&Miglio2013}
{Chaplin}, W.~J. \& {Miglio}, A. 2013, \araa, 51, 353

\bibitem[{{Charpinet} {et~al.}(2005){Charpinet}, {Fontaine}, {Brassard},
  {Green}, \& {Chayer}}]{Charpinet2005}
{Charpinet}, S., {Fontaine}, G., {Brassard}, P., {Green}, E.~M., \& {Chayer},
  P. 2005, \aap, 437, 575

\bibitem[{{Cochran} {et~al.}(1997){Cochran}, {Hatzes}, {Butler}, \&
  {Marcy}}]{Cochran1997}
{Cochran}, W.~D., {Hatzes}, A.~P., {Butler}, R.~P., \& {Marcy}, G.~W. 1997,
  \apj, 483, 457

\bibitem[{{Connelly} {et~al.}(2012){Connelly}, {Bizzarro}, {Krot}, {Nordlund},
  {Wielandt}, \& {Ivanova}}]{Connelly2012}
{Connelly}, J.~N., {Bizzarro}, M., {Krot}, A.~N., {et~al.} 2012, Science, 338,
  651

\bibitem[{{Creevey} {et~al.}(2017){Creevey}, {Metcalfe}, {Schultheis},
  {Salabert}, {Bazot}, {Th{\'e}venin}, {Mathur}, {Xu}, \&
  {Garc{\'\i}a}}]{Creevey2017}
{Creevey}, O.~L., {Metcalfe}, T.~S., {Schultheis}, M., {et~al.} 2017, \aap,
  601, A67

\bibitem[{{Creevey} {et~al.}(2011){Creevey}, {Salabert}, \&
  {Garc{\'\i}a}}]{Creevey2011}
{Creevey}, O.~L., {Salabert}, D., \& {Garc{\'\i}a}, R.~A. 2011, in Journal of
  Physics Conference Series, Vol. 271, GONG-SoHO 24: A New Era of Seismology of
  the Sun and Solar-Like Stars (IOP), 012054

\bibitem[{{Cunha} {et~al.}(2021){Cunha}, {Roxburgh}, {Aguirre B{\o}rsen-Koch},
  {Ball}, {Basu}, {Chaplin}, {Goupil}, {Nsamba}, {Ong}, {Reese}, {Verma},
  {Belkacem}, {Campante}, {Christensen-Dalsgaard}, {Clara}, {Deheuvels},
  {Monteiro}, {Noll}, {Ouazzani}, {R{\o}rsted}, {Stokholm}, \&
  {Winther}}]{Cunha2021}
{Cunha}, M.~S., {Roxburgh}, I.~W., {Aguirre B{\o}rsen-Koch}, V., {et~al.} 2021,
  \mnras, 508, 5864

\bibitem[{{Davies} {et~al.}(2016){Davies}, {Silva Aguirre}, {Bedding}, {Hand
  berg}, {Lund}, {Chaplin}, {Huber}, {White}, {Benomar}, {Hekker}, {Basu},
  {Campante}, {Christensen-Dalsgaard}, {Elsworth}, {Karoff}, {Kjeldsen},
  {Lundkvist}, {Metcalfe}, \& {Stello}}]{Davies2016}
{Davies}, G.~R., {Silva Aguirre}, V., {Bedding}, T.~R., {et~al.} 2016, \mnras,
  456, 2183

\bibitem[{{Di Mauro}(2004)}]{DiMauro2004_theo}
{Di Mauro}, M.~P. 2004, in ESA Special Publication, Vol. 559, SOHO 14 Helio-
  and Asteroseismology: Towards a Golden Future, ed. D.~{Danesy}, 186

\bibitem[{{Efron} \& {Tibshirani}(1993)}]{Efron&Tibshirani1993}
{Efron}, B. \& {Tibshirani}, R.~J. 1993, {An introduction to the bootstrap},
  {Chapman \& Hall/CRC monographs on statistics and applied probability}
  ({London}: {Chapman and Hall})

\bibitem[{{Farnir} {et~al.}(2020){Farnir}, {Dupret}, {Buldgen}, {Salmon},
  {Noels}, {Pin{\c{c}}on}, {Pezzotti}, \& {Eggenberger}}]{Farnir2020}
{Farnir}, M., {Dupret}, M.~A., {Buldgen}, G., {et~al.} 2020, \aap, 644, A37

\bibitem[{{Farnir} {et~al.}(2023){Farnir}, {Valentino}, {Dupret}, \&
  {Broomhall}}]{Farnir2023}
{Farnir}, M., {Valentino}, A., {Dupret}, M.~A., \& {Broomhall}, A.~M. 2023,
  \mnras, 521, 4131

\bibitem[{{Fellay} {et~al.}(2023){Fellay}, {Pezzotti}, {Buldgen},
  {Eggenberger}, \& {Bolmont}}]{Fellay2023}
{Fellay}, L., {Pezzotti}, C., {Buldgen}, G., {Eggenberger}, P., \& {Bolmont},
  E. 2023, \aap, 669, A2

\bibitem[{{Foreman-Mackey} {et~al.}(2013){Foreman-Mackey}, {Hogg}, {Lang}, \&
  {Goodman}}]{Foreman-Mackey2013}
{Foreman-Mackey}, D., {Hogg}, D.~W., {Lang}, D., \& {Goodman}, J. 2013, \pasp,
  125, 306

\bibitem[{{Frandsen} {et~al.}(2002){Frandsen}, {Carrier}, {Aerts}, {Stello},
  {Maas}, {Burnet}, {Bruntt}, {Teixeira}, {de Medeiros}, {Bouchy}, {Kjeldsen},
  {Pijpers}, \& {Christensen-Dalsgaard}}]{Frandsen2002}
{Frandsen}, S., {Carrier}, F., {Aerts}, C., {et~al.} 2002, \aap, 394, L5

\bibitem[{{Furlan} {et~al.}(2018){Furlan}, {Ciardi}, {Cochran}, {Everett},
  {Latham}, {Marcy}, {Buchhave}, {Endl}, {Isaacson}, {Petigura}, {Gautier},
  {Huber}, {Bieryla}, {Borucki}, {Brugamyer}, {Caldwell}, {Cochran}, {Howard},
  {Howell}, {Johnson}, {MacQueen}, {Quinn}, {Robertson}, {Mathur}, \&
  {Batalha}}]{Furlan2018}
{Furlan}, E., {Ciardi}, D.~R., {Cochran}, W.~D., {et~al.} 2018, \apj, 861, 149

\bibitem[{{Gaia Collaboration} {et~al.}(2021){Gaia Collaboration}, {Brown},
  {Vallenari}, {Prusti}, {de Bruijne}, {Babusiaux}, {Biermann}, {Creevey},
  {Evans}, {Eyer}, {Hutton}, {Jansen}, {Jordi}, {Klioner}, {Lammers},
  {Lindegren}, {Luri}, {Mignard}, {Panem}, {Pourbaix}, {Randich}, {Sartoretti},
  {Soubiran}, {Walton}, {Arenou}, {Bailer-Jones}, {Bastian}, {Cropper},
  {Drimmel}, {Katz}, {Lattanzi}, {van Leeuwen}, {Bakker}, {Cacciari},
  {Casta{\~n}eda}, {De Angeli}, {Ducourant}, {Fabricius}, {Fouesneau},
  {Fr{\'e}mat}, {Guerra}, {Guerrier}, {Guiraud}, {Jean-Antoine Piccolo},
  {Masana}, {Messineo}, {Mowlavi}, {Nicolas}, {Nienartowicz}, {Pailler},
  {Panuzzo}, {Riclet}, {Roux}, {Seabroke}, {Sordo}, {Tanga}, {Th{\'e}venin},
  {Gracia-Abril}, {Portell}, {Teyssier}, {Altmann}, {Andrae}, {Bellas-Velidis},
  {Benson}, {Berthier}, {Blomme}, {Brugaletta}, {Burgess}, {Busso}, {Carry},
  {Cellino}, {Cheek}, {Clementini}, {Damerdji}, {Davidson}, {Delchambre},
  {Dell'Oro}, {Fern{\'a}ndez-Hern{\'a}ndez}, {Galluccio}, {Garc{\'\i}a-Lario},
  {Garcia-Reinaldos}, {Gonz{\'a}lez-N{\'u}{\~n}ez}, {Gosset}, {Haigron},
  {Halbwachs}, {Hambly}, {Harrison}, {Hatzidimitriou}, {Heiter},
  {Hern{\'a}ndez}, {Hestroffer}, {Hodgkin}, {Holl}, {Jan{\ss}en}, {Jevardat de
  Fombelle}, {Jordan}, {Krone-Martins}, {Lanzafame}, {L{\"o}ffler}, {Lorca},
  {Manteiga}, {Marchal}, {Marrese}, {Moitinho}, {Mora}, {Muinonen}, {Osborne},
  {Pancino}, {Pauwels}, {Petit}, {Recio-Blanco}, {Richards}, {Riello},
  {Rimoldini}, {Robin}, {Roegiers}, {Rybizki}, {Sarro}, {Siopis}, {Smith},
  {Sozzetti}, {Ulla}, {Utrilla}, {van Leeuwen}, {van Reeven}, {Abbas}, {Abreu
  Aramburu}, {Accart}, {Aerts}, {Aguado}, {Ajaj}, {Altavilla}, {{\'A}lvarez},
  {{\'A}lvarez Cid-Fuentes}, {Alves}, {Anderson}, {Anglada Varela}, {Antoja},
  {Audard}, {Baines}, {Baker}, {Balaguer-N{\'u}{\~n}ez}, {Balbinot}, {Balog},
  {Barache}, {Barbato}, {Barros}, {Barstow}, {Bartolom{\'e}}, {Bassilana},
  {Bauchet}, {Baudesson-Stella}, {Becciani}, {Bellazzini}, {Bernet}, {Bertone},
  {Bianchi}, {Blanco-Cuaresma}, {Boch}, {Bombrun}, {Bossini}, {Bouquillon},
  {Bragaglia}, {Bramante}, {Breedt}, {Bressan}, {Brouillet}, {Bucciarelli},
  {Burlacu}, {Busonero}, {Butkevich}, {Buzzi}, {Caffau}, {Cancelliere},
  {C{\'a}novas}, {Cantat-Gaudin}, {Carballo}, {Carlucci}, {Carnerero},
  {Carrasco}, {Casamiquela}, {Castellani}, {Castro-Ginard}, {Castro Sampol},
  {Chaoul}, {Charlot}, {Chemin}, {Chiavassa}, {Cioni}, {Comoretto}, {Cooper},
  {Cornez}, {Cowell}, {Crifo}, {Crosta}, {Crowley}, {Dafonte}, {Dapergolas},
  {David}, {David}, {de Laverny}, {De Luise}, {De March}, {De Ridder}, {de
  Souza}, {de Teodoro}, {de Torres}, {del Peloso}, {del Pozo}, {Delbo},
  {Delgado}, {Delgado}, {Delisle}, {Di Matteo}, {Diakite}, {Diener},
  {Distefano}, {Dolding}, {Eappachen}, {Edvardsson}, {Enke}, {Esquej}, {Fabre},
  {Fabrizio}, {Faigler}, {Fedorets}, {Fernique}, {Fienga}, {Figueras},
  {Fouron}, {Fragkoudi}, {Fraile}, {Franke}, {Gai}, {Garabato},
  {Garcia-Gutierrez}, {Garc{\'\i}a-Torres}, {Garofalo}, {Gavras}, {Gerlach},
  {Geyer}, {Giacobbe}, {Gilmore}, {Girona}, {Giuffrida}, {Gomel}, {Gomez},
  {Gonzalez-Santamaria}, {Gonz{\'a}lez-Vidal}, {Granvik},
  {Guti{\'e}rrez-S{\'a}nchez}, {Guy}, {Hauser}, {Haywood}, {Helmi}, {Hidalgo},
  {Hilger}, {H{\l}adczuk}, {Hobbs}, {Holland}, {Huckle}, {Jasniewicz},
  {Jonker}, {Juaristi Campillo}, {Julbe}, {Karbevska}, {Kervella}, {Khanna},
  {Kochoska}, {Kontizas}, {Kordopatis}, {Korn}, {Kostrzewa-Rutkowska},
  {Kruszy{\'n}ska}, {Lambert}, {Lanza}, {Lasne}, {Le Campion}, {Le Fustec},
  {Lebreton}, {Lebzelter}, {Leccia}, {Leclerc}, {Lecoeur-Taibi}, {Liao},
  {Licata}, {Lindstr{\o}m}, {Lister}, {Livanou}, {Lobel}, {Madrero Pardo},
  {Managau}, {Mann}, {Marchant}, {Marconi}, {Marcos Santos}, {Marinoni},
  {Marocco}, {Marshall}, {Martin Polo}, {Mart{\'\i}n-Fleitas}, {Masip},
  {Massari}, {Mastrobuono-Battisti}, {Mazeh}, {McMillan}, {Messina},
  {Michalik}, {Millar}, {Mints}, {Molina}, {Molinaro}, {Moln{\'a}r},
  {Montegriffo}, {Mor}, {Morbidelli}, {Morel}, {Morris}, {Mulone}, {Munoz},
  {Muraveva}, {Murphy}, {Musella}, {Noval}, {Ord{\'e}novic}, {Orr{\`u}},
  {Osinde}, {Pagani}, {Pagano}, {Palaversa}, {Palicio}, {Panahi}, {Pawlak},
  {Pe{\~n}alosa Esteller}, {Penttil{\"a}}, {Piersimoni}, {Pineau}, {Plachy},
  {Plum}, {Poggio}, {Poretti}, {Poujoulet}, {Pr{\v{s}}a}, {Pulone}, {Racero},
  {Ragaini}, {Rainer}, {Raiteri}, {Rambaux}, {Ramos}, {Ramos-Lerate}, {Re
  Fiorentin}, {Regibo}, {Reyl{\'e}}, {Ripepi}, {Riva}, {Rixon}, {Robichon},
  {Robin}, {Roelens}, {Rohrbasser}, {Romero-G{\'o}mez}, {Rowell}, {Royer},
  {Rybicki}, {Sadowski}, {Sagrist{\`a} Sell{\'e}s}, {Sahlmann}, {Salgado},
  {Salguero}, {Samaras}, {Sanchez Gimenez}, {Sanna}, {Santove{\~n}a},
  {Sarasso}, {Schultheis}, {Sciacca}, {Segol}, {Segovia}, {S{\'e}gransan},
  {Semeux}, {Shahaf}, {Siddiqui}, {Siebert}, {Siltala}, {Slezak}, {Smart},
  {Solano}, {Solitro}, {Souami}, {Souchay}, {Spagna}, {Spoto}, {Steele},
  {Steidelm{\"u}ller}, {Stephenson}, {S{\"u}veges}, {Szabados}, {Szegedi-Elek},
  {Taris}, {Tauran}, {Taylor}, {Teixeira}, {Thuillot}, {Tonello}, {Torra},
  {Torra}, {Turon}, {Unger}, {Vaillant}, {van Dillen}, {Vanel}, {Vecchiato},
  {Viala}, {Vicente}, {Voutsinas}, {Weiler}, {Wevers}, {Wyrzykowski}, {Yoldas},
  {Yvard}, {Zhao}, {Zorec}, {Zucker}, {Zurbach}, \& {Zwitter}}]{Gaia2021}
{Gaia Collaboration}, {Brown}, A.~G.~A., {Vallenari}, A., {et~al.} 2021, \aap,
  649, A1

\bibitem[{{Garc{\'\i}a} \& {Ballot}(2019)}]{Garcia&Ballot2019}
{Garc{\'\i}a}, R.~A. \& {Ballot}, J. 2019, Living Reviews in Solar Physics, 16,
  4

\bibitem[{{Goupil} {et~al.}(2024){Goupil}, {Catala}, {Samadi}, {Belkacem},
  {Ouazzani}, {Reese}, {Appourchaux}, {Mathur}, {Cabrera}, {B{\"o}rner},
  {Paproth}, {Moedas}, {Verma}, {Lebreton}, {Deal}, {Ballot}, {Chaplin},
  {Christensen-Dalsgaard}, {Cunha}, {Lanza}, {Miglio}, {Morel}, {Serenelli},
  {Mosser}, {Creevey}, {Moya}, {Garcia}, {Nielsen}, \& {Hatt}}]{Goupil2024}
{Goupil}, M.~J., {Catala}, C., {Samadi}, R., {et~al.} 2024, \aap, 683, A78

\bibitem[{{Green} {et~al.}(2018){Green}, {Schlafly}, {Finkbeiner}, {Rix},
  {Martin}, {Burgett}, {Draper}, {Flewelling}, {Hodapp}, {Kaiser}, {Kudritzki},
  {Magnier}, {Metcalfe}, {Tonry}, {Wainscoat}, \& {Waters}}]{Green2018}
{Green}, G.~M., {Schlafly}, E.~F., {Finkbeiner}, D., {et~al.} 2018, \mnras,
  478, 651

\bibitem[{{Grevesse} \& {Sauval}(1998)}]{Grevesse&Sauval1998}
{Grevesse}, N. \& {Sauval}, A.~J. 1998, \ssr, 85, 161

\bibitem[{{Gruberbauer} {et~al.}(2013){Gruberbauer}, {Guenther}, {MacLeod}, \&
  {Kallinger}}]{Gruberbauer2013}
{Gruberbauer}, M., {Guenther}, D.~B., {MacLeod}, K., \& {Kallinger}, T. 2013,
  \mnras, 435, 242

\bibitem[{{Guo} \& {Jiang}(2023)}]{Guo&Jiang2023}
{Guo}, Z. \& {Jiang}, C. 2023, Astronomy and Computing, 42, 100686

\bibitem[{{Hon} {et~al.}(2020){Hon}, {Bellinger}, {Hekker}, {Stello}, \&
  {Kuszlewicz}}]{Hon2020}
{Hon}, M., {Bellinger}, E.~P., {Hekker}, S., {Stello}, D., \& {Kuszlewicz},
  J.~S. 2020, \mnras, 499, 2445

\bibitem[{{Howell} {et~al.}(2012){Howell}, {Rowe}, {Bryson}, {Quinn}, {Marcy},
  {Isaacson}, {Ciardi}, {Chaplin}, {Metcalfe}, {Monteiro}, {Appourchaux},
  {Basu}, {Creevey}, {Gilliland}, {Quirion}, {Stello}, {Kjeldsen},
  {Christensen-Dalsgaard}, {Elsworth}, {Garc{\'\i}a}, {Houdek}, {Karoff},
  {Molenda-{\.Z}akowicz}, {Thompson}, {Verner}, {Torres}, {Fressin}, {Crepp},
  {Adams}, {Dupree}, {Sasselov}, {Dressing}, {Borucki}, {Koch}, {Lissauer},
  {Latham}, {Buchhave}, {Gautier}, {Everett}, {Horch}, {Batalha}, {Dunham},
  {Szkody}, {Silva}, {Mighell}, {Holberg}, {Ballot}, {Bedding}, {Bruntt},
  {Campante}, {Handberg}, {Hekker}, {Huber}, {Mathur}, {Mosser}, {R{\'e}gulo},
  {White}, {Christiansen}, {Middour}, {Haas}, {Hall}, {Jenkins}, {McCaulif},
  {Fanelli}, {Kulesa}, {McCarthy}, \& {Henze}}]{Howell2012}
{Howell}, S.~B., {Rowe}, J.~F., {Bryson}, S.~T., {et~al.} 2012, \apj, 746, 123

\bibitem[{{Howell} {et~al.}(2014){Howell}, {Sobeck}, {Haas}, {Still},
  {Barclay}, {Mullally}, {Troeltzsch}, {Aigrain}, {Bryson}, {Caldwell},
  {Chaplin}, {Cochran}, {Huber}, {Marcy}, {Miglio}, {Najita}, {Smith},
  {Twicken}, \& {Fortney}}]{Howell2014}
{Howell}, S.~B., {Sobeck}, C., {Haas}, M., {et~al.} 2014, \pasp, 126, 398

\bibitem[{{Jiang} \& {Gizon}(2021)}]{Jiang&Gizon2021}
{Jiang}, C. \& {Gizon}, L. 2021, Research in Astronomy and Astrophysics, 21,
  226

\bibitem[{{J{\o}rgensen} {et~al.}(2021){J{\o}rgensen}, {Montalb{\'a}n},
  {Angelou}, {Miglio}, {Weiss}, {Scuflaire}, {Noels}, {Mosumgaard}, \& {Silva
  Aguirre}}]{Jorgensen2021}
{J{\o}rgensen}, A. C.~S., {Montalb{\'a}n}, J., {Angelou}, G.~C., {et~al.} 2021,
  \mnras, 500, 4277

\bibitem[{{J{\o}rgensen} {et~al.}(2020){J{\o}rgensen}, {Montalb{\'a}n},
  {Miglio}, {Rendle}, {Davies}, {Buldgen}, {Scuflaire}, {Noels}, {Gaulme}, \&
  {Garc{\'\i}a}}]{Jorgensen2020}
{J{\o}rgensen}, A. C.~S., {Montalb{\'a}n}, J., {Miglio}, A., {et~al.} 2020,
  \mnras, 495, 4965

\bibitem[{{Kippenhahn} {et~al.}(2012){Kippenhahn}, {Weigert}, \&
  {Weiss}}]{Kippenhahn2012}
{Kippenhahn}, R., {Weigert}, A., \& {Weiss}, A. 2012, {Stellar Structure and
  Evolution}, Astronomy and Astrophysics Library (Springer Berlin Heidelberg)

\bibitem[{{Kirkpatrick}(1984)}]{Kirkpatrick1984}
{Kirkpatrick}, S. 1984, Journal of Statistical Physics, 34, 975

\bibitem[{{Kirkpatrick} {et~al.}(1983){Kirkpatrick}, {Gelatt}, \&
  {Vecchi}}]{Kirkpatrick1983}
{Kirkpatrick}, S., {Gelatt}, C.~D., \& {Vecchi}, M.~P. 1983, Science, 220, 671

\bibitem[{{Kjeldsen} {et~al.}(2008){Kjeldsen}, {Bedding}, \&
  {Christensen-Dalsgaard}}]{Kjeldsen2008}
{Kjeldsen}, H., {Bedding}, T.~R., \& {Christensen-Dalsgaard}, J. 2008, \apjl,
  683, L175

\bibitem[{{Kosovichev} \& {Kitiashvili}(2020)}]{Kosovichev&Kitiashvili2020}
{Kosovichev}, A.~G. \& {Kitiashvili}, I.~N. 2020, in Solar and Stellar Magnetic
  Fields: Origins and Manifestations, ed. A.~{Kosovichev}, S.~{Strassmeier}, \&
  M.~{Jardine}, Vol. 354, 107--115

\bibitem[{{Lanza}(2015)}]{Lanza2015}
{Lanza}, A.~F. 2015, in Cambridge Workshop on Cool Stars, Stellar Systems, and
  the Sun, Vol.~18, 18th Cambridge Workshop on Cool Stars, Stellar Systems, and
  the Sun, 811--830

\bibitem[{{Lebreton} {et~al.}(2014){Lebreton}, {Goupil}, \&
  {Montalb{\'a}n}}]{Lebreton2014}
{Lebreton}, Y., {Goupil}, M.~J., \& {Montalb{\'a}n}, J. 2014, in EAS
  Publications Series, Vol.~65, EAS Publications Series, ed. Y.~{Lebreton},
  D.~{Valls-Gabaud}, \& C.~{Charbonnel} (EDP), 99--176

\bibitem[{{Lindegren} {et~al.}(2021){Lindegren}, {Bastian}, {Biermann},
  {Bombrun}, {de Torres}, {Gerlach}, {Geyer}, {Hern{\'a}ndez}, {Hilger},
  {Hobbs}, {Klioner}, {Lammers}, {McMillan}, {Ramos-Lerate},
  {Steidelm{\"u}ller}, {Stephenson}, \& {van Leeuwen}}]{Lindegren2021}
{Lindegren}, L., {Bastian}, U., {Biermann}, M., {et~al.} 2021, \aap, 649, A4

\bibitem[{{Lund} {et~al.}(2017){Lund}, {Silva Aguirre}, {Davies}, {Chaplin},
  {Christensen-Dalsgaard}, {Houdek}, {White}, {Bedding}, {Ball}, {Huber},
  {Antia}, {Lebreton}, {Latham}, {Handberg}, {Verma}, {Basu}, {Casagrande},
  {Justesen}, {Kjeldsen}, \& {Mosumgaard}}]{Lund2017}
{Lund}, M.~N., {Silva Aguirre}, V., {Davies}, G.~R., {et~al.} 2017, \apj, 835,
  172

\bibitem[{{Maeder}(2009)}]{Maeder2009}
{Maeder}, A. 2009, {Physics, Formation and Evolution of Rotating Stars}
  (Springer Verlag)

\bibitem[{{Marcy} {et~al.}(2014){Marcy}, {Isaacson}, {Howard}, {Rowe},
  {Jenkins}, {Bryson}, {Latham}, {Howell}, {Gautier}, {Batalha}, {Rogers},
  {Ciardi}, {Fischer}, {Gilliland}, {Kjeldsen}, {Christensen-Dalsgaard},
  {Huber}, {Chaplin}, {Basu}, {Buchhave}, {Quinn}, {Borucki}, {Koch}, {Hunter},
  {Caldwell}, {Van Cleve}, {Kolbl}, {Weiss}, {Petigura}, {Seager}, {Morton},
  {Johnson}, {Ballard}, {Burke}, {Cochran}, {Endl}, {MacQueen}, {Everett},
  {Lissauer}, {Ford}, {Torres}, {Fressin}, {Brown}, {Steffen}, {Charbonneau},
  {Basri}, {Sasselov}, {Winn}, {Sanchis-Ojeda}, {Christiansen}, {Adams},
  {Henze}, {Dupree}, {Fabrycky}, {Fortney}, {Tarter}, {Holman}, {Tenenbaum},
  {Shporer}, {Lucas}, {Welsh}, {Orosz}, {Bedding}, {Campante}, {Davies},
  {Elsworth}, {Handberg}, {Hekker}, {Karoff}, {Kawaler}, {Lund}, {Lundkvist},
  {Metcalfe}, {Miglio}, {Silva Aguirre}, {Stello}, {White}, {Boss}, {Devore},
  {Gould}, {Prsa}, {Agol}, {Barclay}, {Coughlin}, {Brugamyer}, {Mullally},
  {Quintana}, {Still}, {Thompson}, {Morrison}, {Twicken}, {D{\'e}sert},
  {Carter}, {Crepp}, {H{\'e}brard}, {Santerne}, {Moutou}, {Sobeck}, {Hudgins},
  {Haas}, {Robertson}, {Lillo-Box}, \& {Barrado}}]{Marcy2014}
{Marcy}, G.~W., {Isaacson}, H., {Howard}, A.~W., {et~al.} 2014, \apjs, 210, 20

\bibitem[{{Metcalfe} \& {Charbonneau}(2003)}]{Metcalfe&Charbonneau2003}
{Metcalfe}, T.~S. \& {Charbonneau}, P. 2003, Journal of Computational Physics,
  185, 176

\bibitem[{{Metcalfe} {et~al.}(2009){Metcalfe}, {Creevey}, \&
  {Christensen-Dalsgaard}}]{Metcalfe2009}
{Metcalfe}, T.~S., {Creevey}, O.~L., \& {Christensen-Dalsgaard}, J. 2009, \apj,
  699, 373

\bibitem[{{Metcalfe} {et~al.}(2014){Metcalfe}, {Creevey}, {Do{\u{g}}an},
  {Mathur}, {Xu}, {Bedding}, {Chaplin}, {Christensen-Dalsgaard}, {Karoff},
  {Trampedach}, {Benomar}, {Brown}, {Buzasi}, {Campante}, {{\c{C}}elik},
  {Cunha}, {Davies}, {Deheuvels}, {Derekas}, {Di Mauro}, {Garc{\'\i}a},
  {Guzik}, {Howe}, {MacGregor}, {Mazumdar}, {Montalb{\'a}n}, {Monteiro},
  {Salabert}, {Serenelli}, {Stello}, {Ste\&{\c{s}}acute}, {licki}, {Suran},
  {Y{\i}ld{\i}z}, {Aksoy}, {Elsworth}, {Gruberbauer}, {Guenther}, {Lebreton},
  {Molaverdikhani}, {Pricopi}, {Simoniello}, \& {White}}]{Metcalfe2014}
{Metcalfe}, T.~S., {Creevey}, O.~L., {Do{\u{g}}an}, G., {et~al.} 2014, \apjs,
  214, 27

\bibitem[{{Metcalfe} {et~al.}(2023){Metcalfe}, {Townsend}, \&
  {Ball}}]{Metcalfe2023}
{Metcalfe}, T.~S., {Townsend}, R. H.~D., \& {Ball}, W.~H. 2023, Research Notes
  of the American Astronomical Society, 7, 164

\bibitem[{{Miglio} {et~al.}(2021){Miglio}, {Girardi}, {Grundahl}, {Mosser},
  {Bastian}, {Bragaglia}, {Brogaard}, {Buldgen}, {Chantereau}, {Chaplin},
  {Chiappini}, {Dupret}, {Eggenberger}, {Gieles}, {Izzard}, {Kawata}, {Karoff},
  {Lagarde}, {Mackereth}, {Magrin}, {Meynet}, {Michel}, {Montalb{\'a}n},
  {Nascimbeni}, {Noels}, {Piotto}, {Ragazzoni}, {Soszy{\'n}ski}, {Tolstoy},
  {Toonen}, {Triaud}, \& {Vincenzo}}]{Miglio2021}
{Miglio}, A., {Girardi}, L., {Grundahl}, F., {et~al.} 2021, Experimental
  Astronomy, 51, 963

\bibitem[{{Miglio} \& {Montalb{\'a}n}(2005)}]{Miglio&Montalban2005}
{Miglio}, A. \& {Montalb{\'a}n}, J. 2005, \aap, 441, 615

\bibitem[{{Nagayama} {et~al.}(2019){Nagayama}, {Bailey}, {Loisel}, {Dunham},
  {Rochau}, {Blancard}, {Colgan}, {Coss{\'e}}, {Faussurier}, {Fontes},
  {Gilleron}, {Hansen}, {Iglesias}, {Golovkin}, {Kilcrease}, {MacFarlane},
  {Mancini}, {More}, {Orban}, {Pain}, {Sherrill}, \& {Wilson}}]{Nagayama2019}
{Nagayama}, T., {Bailey}, J.~E., {Loisel}, G.~P., {et~al.} 2019, \prl, 122,
  235001

\bibitem[{{Nordlund} \& {Stein}(1997)}]{Nordlund&Stein1997}
{Nordlund}, A. \& {Stein}, R. 1997, in Astrophysics and Space Science Library,
  Vol. 225, SCORe'96 : Solar Convection and Oscillations and their
  Relationship, ed. F.~P. {Pijpers}, J.~{Christensen-Dalsgaard}, \& C.~S.
  {Rosenthal}, 79--103

\bibitem[{{Nsamba} {et~al.}(2018){Nsamba}, {Campante}, {Monteiro}, {Cunha},
  {Rendle}, {Reese}, \& {Verma}}]{Nsamba2018}
{Nsamba}, B., {Campante}, T.~L., {Monteiro}, M.~J.~P.~F.~G., {et~al.} 2018,
  \mnras, 477, 5052

\bibitem[{{Nsamba} {et~al.}(2022){Nsamba}, {Cunha}, {Rocha}, {Pereira},
  {Monteiro}, \& {Campante}}]{Nsamba2022}
{Nsamba}, B., {Cunha}, M.~S., {Rocha}, C. I.~S.~A., {et~al.} 2022, \mnras, 514,
  893

\bibitem[{{Ot{\'\i} Floranes} {et~al.}(2005){Ot{\'\i} Floranes},
  {Christensen-Dalsgaard}, \& {Thompson}}]{Oti2005}
{Ot{\'\i} Floranes}, H., {Christensen-Dalsgaard}, J., \& {Thompson}, M.~J.
  2005, \mnras, 356, 671

\bibitem[{Pedregosa {et~al.}(2011)Pedregosa, Varoquaux, Gramfort, Michel,
  Thirion, Grisel, Blondel, Prettenhofer, Weiss, Dubourg, Vanderplas, Passos,
  Cournapeau, Brucher, Perrot, \& Duchesnay}]{Scikit-learn2011}
Pedregosa, F., Varoquaux, G., Gramfort, A., {et~al.} 2011, Journal of Machine
  Learning Research, 12, 2825

\bibitem[{{P{\'e}rez Hern{\'a}ndez} {et~al.}(2019){P{\'e}rez Hern{\'a}ndez},
  {Garc{\'\i}a}, {Mathur}, {Santos}, \& {R{\'e}gulo}}]{PerezHernandez2019}
{P{\'e}rez Hern{\'a}ndez}, F., {Garc{\'\i}a}, R.~A., {Mathur}, S., {Santos}, A.
  R.~G., \& {R{\'e}gulo}, C. 2019, Frontiers in Astronomy and Space Sciences,
  6, 41

\bibitem[{{Pezzotti} {et~al.}(2026){Pezzotti}, {B{\'e}trisey}, {Buldgen},
  {Gilfanov}, {Bikmaev}, {Sunyaev}, {I{\textcommabelow s}{\i}k}, {Gosset}, \&
  {Wright}}]{Pezzotti2026}
{Pezzotti}, C., {B{\'e}trisey}, J., {Buldgen}, G., {et~al.} 2026, \aap, 706,
  A257

\bibitem[{{Pezzotti} {et~al.}(2021){Pezzotti}, {Eggenberger}, {Buldgen},
  {Meynet}, {Bourrier}, \& {Mordasini}}]{Pezzotti2021}
{Pezzotti}, C., {Eggenberger}, P., {Buldgen}, G., {et~al.} 2021, \aap, 650,
  A108

\bibitem[{{Pezzotti} {et~al.}(2022){Pezzotti}, {Ottoni}, {Buldgen}, {Lyttle},
  {Eggenberger}, {Udry}, {S{\'e}gransan}, {Mayor}, {Lovis}, {Marmier},
  {Miglio}, {Elsworth}, {Davies}, \& {Ball}}]{Pezzotti2022}
{Pezzotti}, C., {Ottoni}, G., {Buldgen}, G., {et~al.} 2022, \aap, 657, A89

\bibitem[{{Pinsonneault} {et~al.}(2012){Pinsonneault}, {An},
  {Molenda-{\.Z}akowicz}, {Chaplin}, {Metcalfe}, \&
  {Bruntt}}]{Pinsonneault2012}
{Pinsonneault}, M.~H., {An}, D., {Molenda-{\.Z}akowicz}, J., {et~al.} 2012,
  \apjs, 199, 30

\bibitem[{{Press} {et~al.}(1986){Press}, {Flannery}, \&
  {Teukolsky}}]{Press1986}
{Press}, W.~H., {Flannery}, B.~P., \& {Teukolsky}, S.~A. 1986, {Numerical
  recipes. The art of scientific computing} (Cambridge: University Press)

\bibitem[{{Privitera} {et~al.}(2016{\natexlab{a}}){Privitera}, {Meynet},
  {Eggenberger}, {Georgy}, {Ekstr{\"o}m}, {Vidotto}, {Bianda}, {Villaver}, \&
  {ud-Doula}}]{Privitera2016_RG}
{Privitera}, G., {Meynet}, G., {Eggenberger}, P., {et~al.} 2016{\natexlab{a}},
  \aap, 593, L15

\bibitem[{{Privitera} {et~al.}(2016{\natexlab{b}}){Privitera}, {Meynet},
  {Eggenberger}, {Vidotto}, {Villaver}, \& {Bianda}}]{Privitera2016_spi1}
{Privitera}, G., {Meynet}, G., {Eggenberger}, P., {et~al.} 2016{\natexlab{b}},
  \aap, 591, A45

\bibitem[{{Privitera} {et~al.}(2016{\natexlab{c}}){Privitera}, {Meynet},
  {Eggenberger}, {Vidotto}, {Villaver}, \& {Bianda}}]{Privitera2016_spi2}
{Privitera}, G., {Meynet}, G., {Eggenberger}, P., {et~al.} 2016{\natexlab{c}},
  \aap, 593, A128

\bibitem[{{Rao} {et~al.}(2018){Rao}, {Meynet}, {Eggenberger}, {Haemmerl{\'e}},
  {Privitera}, {Georgy}, {Ekstr{\"o}m}, \& {Mordasini}}]{Rao2018}
{Rao}, S., {Meynet}, G., {Eggenberger}, P., {et~al.} 2018, \aap, 618, A18

\bibitem[{{Rauer} {et~al.}(2025){Rauer}, {Aerts}, {Cabrera}, {Deleuil},
  {Erikson}, {Gizon}, {Goupil}, {Heras}, {Walloschek}, {Lorenzo-Alvarez},
  {Marliani}, {Martin-Garcia}, {Mas-Hesse}, {O'Rourke}, {Osborn}, {Pagano},
  {Piotto}, {Pollacco}, {Ragazzoni}, {Ramsay}, {Udry}, {Appourchaux}, {Benz},
  {Brandeker}, {G{\"u}del}, {Janot-Pacheco}, {Kabath}, {Kjeldsen}, {Min},
  {Santos}, {Smith}, {Suarez}, {Werner}, {Aboudan}, {Abreu}, {Acu{\~n}a},
  {Adams}, {Adibekyan}, {Affer}, {Agneray}, {Agnor}, {Aguirre B{\o}rsen-Koch},
  {Ahmed}, {Aigrain}, {Al-Bahlawan}, {Alcacera Gil}, {Alei}, {Alencar},
  {Alexander}, {Alfonso-Garz{\'o}n}, {Alibert}, {Allende Prieto}, {Almeida},
  {Alonso Sobrino}, {Altavilla}, {Althaus}, {Alvarez Trujillo}, {Amarsi},
  {Ammler-von Eiff}, {Am{\^o}res}, {Andrade}, {Antoniadis-Karnavas},
  {Ant{\'o}nio}, {Aparicio del Moral}, {Appolloni}, {Arena}, {Armstrong},
  {Aroca Aliaga}, {Asplund}, {Audenaert}, {Auricchio}, {Avelino}, {Baeke},
  {Bailli{\'e}}, {Balado}, {Ballber Balaguer{\'o}}, {Balestra}, {Ball},
  {Ballans}, {Ballot}, {Barban}, {Barbary}, {Barbieri}, {Barcel{\'o} Forteza},
  {Barker}, {Barklem}, {Barnes}, {Barrado Navascues}, {Barragan}, {Baruteau},
  {Basu}, {Baudin}, {Baumeister}, {Bayliss}, {Bazot}, {Beck}, {Belkacem},
  {Bellinger}, {Benatti}, {Benomar}, {B{\'e}rard}, {Bergemann}, {Bergomi},
  {Bernardo}, {Biazzo}, {Bignamini}, {Bigot}, {Billot}, {Binet}, {Biondi},
  {Biondi}, {Birch}, {Bitsch}, {Bluhm Ceballos}, {B{\'o}di}, {Bogn{\'a}r},
  {Boisse}, {Bolmont}, {Bonanno}, {Bonavita}, {Bonfanti}, {Bonfils}, {Bonito},
  {Bonomo}, {B{\"o}rner}, {Boro Saikia}, {Borreguero Mart{\'\i}n}, {Borsa},
  {Borsato}, {Bossini}, {Bouchy}, {Bou{\'e}}, {Boufleur}, {Boumier},
  {Bourrier}, {Bowman}, {Bozzo}, {Bradley}, {Bray}, {Bressan}, {Breton},
  {Brienza}, {Brito}, {Brogi}, {Brown}, {Brown}, {Brun}, {Bruno}, {Bruns},
  {Buchhave}, {Bugnet}, {Buldgen}, {Burgess}, {Busatta}, {Busso}, {Buzasi},
  {Caballero}, {Cabral}, {Cabrero Gomez}, {Calderone}, {Cameron}, {Cameron},
  {Campante}, {Campos Gestal}, {Canto Martins}, {Cara}, {Carone}, {Carrasco},
  {Casagrande}, {Casewell}, {Cassisi}, {Castellani}, {Castro}, {Catala},
  {Catal{\'a}n Fern{\'a}ndez}, {Catelan}, {Cegla}, {Cerruti}, {Cessa},
  {Chadid}, {Chaplin}, {Charpinet}, {Chiappini}, {Chiarucci}, {Chiavassa},
  {Chinellato}, {Chirulli}, {Christensen-Dalsgaard}, {Church}, {Claret},
  {Clarke}, {Claudi}, {Clermont}, {Coelho}, {Coelho}, {Cogato}, {Colom{\'e}},
  {Condamin}, {Conde Garc{\'\i}a}, \& {Conseil}}]{Rauer2025}
{Rauer}, H., {Aerts}, C., {Cabrera}, J., {et~al.} 2025, Experimental Astronomy,
  59, 26

\bibitem[{{Reese} {et~al.}(2012){Reese}, {Marques}, {Goupil}, {Thompson}, \&
  {Deheuvels}}]{Reese2012}
{Reese}, D.~R., {Marques}, J.~P., {Goupil}, M.~J., {Thompson}, M.~J., \&
  {Deheuvels}, S. 2012, \aap, 539, A63

\bibitem[{{Rendle} {et~al.}(2019){Rendle}, {Buldgen}, {Miglio}, {Reese},
  {Noels}, {Davies}, {Campante}, {Chaplin}, {Lund}, {Kuszlewicz}, {Scott},
  {Scuflaire}, {Ball}, {Smetana}, \& {Nsamba}}]{Rendle2019}
{Rendle}, B.~M., {Buldgen}, G., {Miglio}, A., {et~al.} 2019, \mnras, 484, 771

\bibitem[{{Ricker} {et~al.}(2015){Ricker}, {Winn}, {Vanderspek}, {Latham},
  {Bakos}, {Bean}, {Berta-Thompson}, {Brown}, {Buchhave}, {Butler}, {Butler},
  {Chaplin}, {Charbonneau}, {Christensen-Dalsgaard}, {Clampin}, {Deming},
  {Doty}, {De Lee}, {Dressing}, {Dunham}, {Endl}, {Fressin}, {Ge}, {Henning},
  {Holman}, {Howard}, {Ida}, {Jenkins}, {Jernigan}, {Johnson}, {Kaltenegger},
  {Kawai}, {Kjeldsen}, {Laughlin}, {Levine}, {Lin}, {Lissauer}, {MacQueen},
  {Marcy}, {McCullough}, {Morton}, {Narita}, {Paegert}, {Palle}, {Pepe},
  {Pepper}, {Quirrenbach}, {Rinehart}, {Sasselov}, {Sato}, {Seager},
  {Sozzetti}, {Stassun}, {Sullivan}, {Szentgyorgyi}, {Torres}, {Udry}, \&
  {Villasenor}}]{Ricker2015}
{Ricker}, G.~R., {Winn}, J.~N., {Vanderspek}, R., {et~al.} 2015, Journal of
  Astronomical Telescopes, Instruments, and Systems, 1, 014003

\bibitem[{{Roxburgh}(2016)}]{Roxburgh2016}
{Roxburgh}, I.~W. 2016, \aap, 585, A63

\bibitem[{{Roxburgh} \& {Vorontsov}(2003)}]{Roxburgh&Vorontsov2003}
{Roxburgh}, I.~W. \& {Vorontsov}, S.~V. 2003, \aap, 411, 215

\bibitem[{{Salaris} \& {Cassisi}(2005)}]{Salaris&Cassisi2005}
{Salaris}, M. \& {Cassisi}, S. 2005, {Evolution of Stars and Stellar
  Populations} (John Wiley \& Sons, Ltd)

\bibitem[{{Salmon} {et~al.}(2021){Salmon}, {Van Grootel}, {Buldgen}, {Dupret},
  \& {Eggenberger}}]{Salmon2021}
{Salmon}, S.~J.~A.~J., {Van Grootel}, V., {Buldgen}, G., {Dupret}, M.~A., \&
  {Eggenberger}, P. 2021, \aap, 646, A7

\bibitem[{{Shibahashi}(1979)}]{Shibahashi1979}
{Shibahashi}, H. 1979, \pasj, 31, 87

\bibitem[{{Silva Aguirre} {et~al.}(2015){Silva Aguirre}, {Davies}, {Basu},
  {Christensen-Dalsgaard}, {Creevey}, {Metcalfe}, {Bedding}, {Casagrande},
  {Handberg}, {Lund}, {Nissen}, {Chaplin}, {Huber}, {Serenelli}, {Stello}, {Van
  Eylen}, {Campante}, {Elsworth}, {Gilliland}, {Hekker}, {Karoff}, {Kawaler},
  {Kjeldsen}, \& {Lundkvist}}]{Silva-Aguirre2015}
{Silva Aguirre}, V., {Davies}, G.~R., {Basu}, S., {et~al.} 2015, \mnras, 452,
  2127

\bibitem[{{Silva Aguirre} {et~al.}(2017){Silva Aguirre}, {Lund}, {Antia},
  {Ball}, {Basu}, {Christensen-Dalsgaard}, {Lebreton}, {Reese}, {Verma},
  {Casagrande}, {Justesen}, {Mosumgaard}, {Chaplin}, {Bedding}, {Davies},
  {Handberg}, {Houdek}, {Huber}, {Kjeldsen}, {Latham}, {White}, {Coelho},
  {Miglio}, \& {Rendle}}]{Silva-Aguirre2017}
{Silva Aguirre}, V., {Lund}, M.~N., {Antia}, H.~M., {et~al.} 2017, \apj, 835,
  173

\bibitem[{{Sonoi} {et~al.}(2015){Sonoi}, {Samadi}, {Belkacem}, {Ludwig},
  {Caffau}, \& {Mosser}}]{Sonoi2015}
{Sonoi}, T., {Samadi}, R., {Belkacem}, K., {et~al.} 2015, \aap, 583, A112

\bibitem[{{Stein} \& {Nordlund}(1998)}]{Stein&Nordlund1998}
{Stein}, R.~F. \& {Nordlund}, {\r{A}}. 1998, \apj, 499, 914

\bibitem[{{Tanner} {et~al.}(2013){Tanner}, {Basu}, \& {Demarque}}]{Tanner2013}
{Tanner}, J.~D., {Basu}, S., \& {Demarque}, P. 2013, \apj, 767, 78

\bibitem[{{Tassoul}(1980)}]{Tassoul1980}
{Tassoul}, M. 1980, \apjs, 43, 469

\bibitem[{{Teixeira} {et~al.}(2003){Teixeira}, {Christensen-Dalsgaard},
  {Carrier}, {Aerts}, {Frandsen}, {Stello}, {Maas}, {Burnet}, {Bruntt}, {de
  Medeiros}, {Bouchy}, {Kjeldsen}, \& {Pijpers}}]{Teixeira2003}
{Teixeira}, T.~C., {Christensen-Dalsgaard}, J., {Carrier}, F., {et~al.} 2003,
  \apss, 284, 233

\bibitem[{{Thomas} {et~al.}(2021){Thomas}, {Chaplin}, {Basu}, {Rendle},
  {Davies}, \& {Miglio}}]{Thomas2021}
{Thomas}, A. E.~L., {Chaplin}, W.~J., {Basu}, S., {et~al.} 2021, \mnras, 502,
  5808

\bibitem[{{Vandakurov}(1967)}]{Vandakurov1967}
{Vandakurov}, Y.~V. 1967, \azh, 44, 786

\bibitem[{{Verma} \& {Silva Aguirre}(2019)}]{Verma2019}
{Verma}, K. \& {Silva Aguirre}, V. 2019, \mnras, 489, 1850

\bibitem[{{Vidotto}(2020)}]{Vidotto2020}
{Vidotto}, A.~A. 2020, in Solar and Stellar Magnetic Fields: Origins and
  Manifestations, ed. A.~{Kosovichev}, S.~{Strassmeier}, \& M.~{Jardine}, Vol.
  354, 259--267

\bibitem[{{White} {et~al.}(2011){White}, {Bedding}, {Stello},
  {Christensen-Dalsgaard}, {Huber}, \& {Kjeldsen}}]{White2011}
{White}, T.~R., {Bedding}, T.~R., {Stello}, D., {et~al.} 2011, \apj, 743, 161

\end{thebibliography}
\enlargethispage{1\baselineskip}

\appendix

\section{Upgraded Spelaion grid}
\label{sec:app:upgraded_spelaion_grid}
We summarised in Tables~\ref{tab:Spelaion_statistics} and~\ref{tab:Spelaion_properties} the statistics and mesh properties of the Spelaion grid and its subgrids. The first row of Table~\ref{tab:Spelaion_statistics} reports statistics for unique models only, as overlaps exist between subgrids.

\begin{table}[h!]
\centering
\caption{Statistics of Spelaion and its subgrids.}
\begin{tabular}{lccc}
\hline 
 & Tracks & Models & Modes \\
\hline \hline
\emph{Spelaion} & \num{35267} & \num{6722883} & 653 million \\ 
Metallic Sun & 1176 &  \num{149398} & 18 million \\  
Standard MS & \num{10192} & \num{1682934} & 160 million \\  
Low-metallicity MS & 8580 & \num{1224144} & 113 million \\
Overshooting MS & \num{26390} & \num{5561380} & 537\ million \\
\hline 
\end{tabular}
\label{tab:Spelaion_statistics}
\end{table}

\begin{table}[h!]
\centering
\caption{Mesh properties of the Spelaion subgrids.}
\begin{tabular}{lccc}
\hline 
 &  Minimum & Maximum & Step \\ 
\hline \hline 
\textit{Metallic Sun} &  &  &  \\ 
Mass $(M_\odot)$ & 0.94 & 1.06 & 0.02 \\ 
$X_0$ & 0.64 & 0.71 & 0.01 \\ 
$Z_0$ & 0.020 & 0.040 & 0.001 \\  
Overshooting & \multicolumn{3}{c}{$\rm\alpha_{ov}=0.00$} \\ 
\hline 
\textit{Standard MS} &  &  &  \\ 
Mass $(M_\odot)$ & 0.70 & 1.66 & 0.02 \\  
$X_0$ & 0.67 & 0.74 & 0.01 \\  
$Z_0$ & 0.005 & 0.030 & 0.001 \\ 
Overshooting & \multicolumn{3}{c}{$\rm\alpha_{ov}=0.00$} \\ 
\hline 
\textit{Low-metallicity MS} &  &  &  \\ 
Mass $(M_\odot)$ & 0.70 & 1.20 & 0.02 \\  
$X_0$ & 0.67 & 0.77 & 0.01 \\  
$Z_0$ & 0.001 & 0.030 & 0.001 \\ 
Overshooting & \multicolumn{3}{c}{$\rm\alpha_{ov}=0.00$} \\
\hline
\textit{Overshooting MS} &  &  &  \\  
Mass $(M_\odot)$ & 1.10 & 1.66 & 0.02 \\  
$X_0$ & 0.68 & 0.74 & 0.01 \\  
$Z_0$ & 0.005 & 0.030 & 0.001 \\ 
Overshooting & 0.00 & 0.20 & 0.05 \\ 
\hline 
\end{tabular}
\label{tab:Spelaion_properties}
\end{table}

\section{Supplementary data}
\label{sec:app:supplementary_data}
\underline{\textit{Empirical uncertainty relationships}}
\vspace*{-0.5ex}

\begin{figure}[t!]
\centering
\begin{subfigure}[b]{.43\textwidth}
  \includegraphics[width=.99\textwidth]{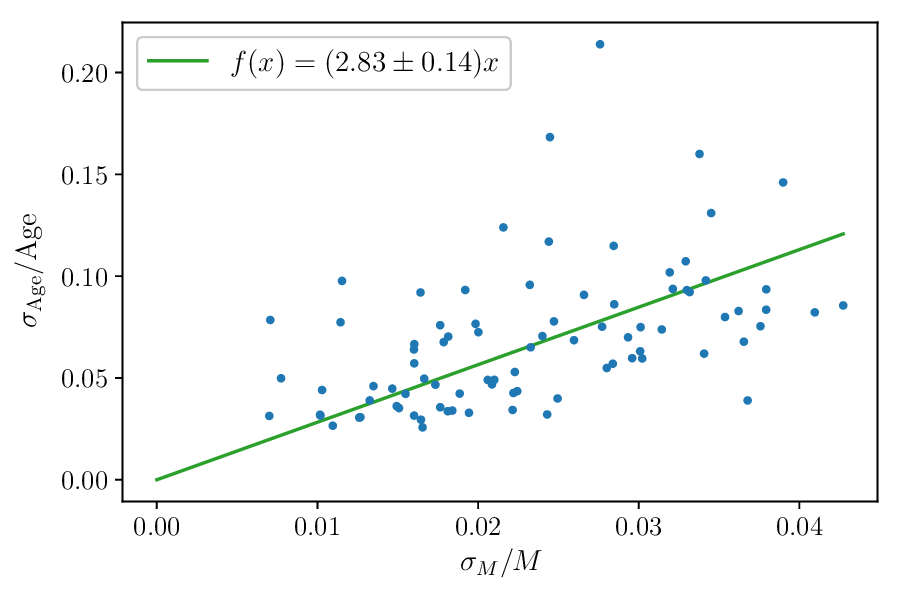}  
\end{subfigure}
\begin{subfigure}[b]{.43\textwidth}
  \includegraphics[width=.99\textwidth]{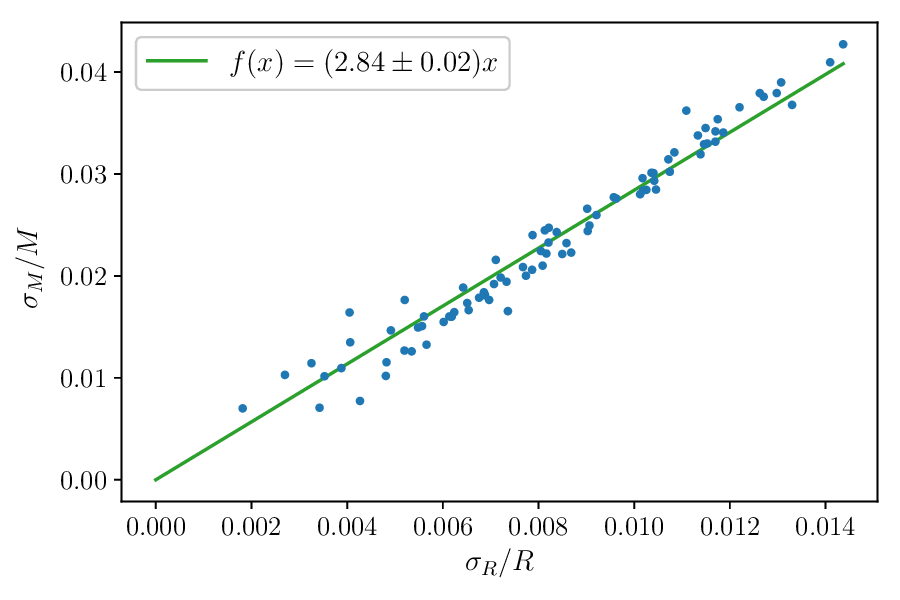}  
\end{subfigure}
\caption{Scaling relations between age and mass relative uncertainties (upper panel), and mass and radius relative uncertainties (lower panel). The green line represent the linear regression with an offset fixed at zero.}
\label{fig:slope_error}
\end{figure}

Figure~\ref{fig:slope_error} illustrates the empirical scaling relations governing the relative uncertainties in stellar mass, radius, and age. It is important to note that these `relations' should be regarded as rules of thumb rather than physical relations, as previous studies have reported scaling factors ranging from 2 to 4. In our analysis, we derive a factor of approximately 2.8 for both the age-mass and mass-radius uncertainty planes, which aligns well with empirical expectations.

Notably, the age-mass relation exhibits substantially greater scatter compared to the mass-radius relation. This discrepancy arises because the stellar radius is more tightly constrained by seismic frequencies. Specifically, individual oscillation frequencies scale with the square root of the mean stellar density \citep[e.g.][]{Vandakurov1967,Shibahashi1979,Tassoul1980}, thereby providing a more direct constraint on the radius. Additionally, mass and age are optimised parameters in our modelling framework, while radius is derived from the optimised parameters. As pointed out in Sect.~\ref{sec:comparison_SA15}, the \citet{Davies2016} sample has a significantly lower statistics than the LEGACY sample and an overall lower data-quality. The mirror plots therefore show a larger scattering. 

\underline{\textit{Comparison with Silva-Aguirre et al. (2015)}}
\vspace*{-0.5ex}

\begin{figure*}[t!]
\centering
\begin{subfigure}[b]{.32\textwidth}
  \includegraphics[width=.99\textwidth]{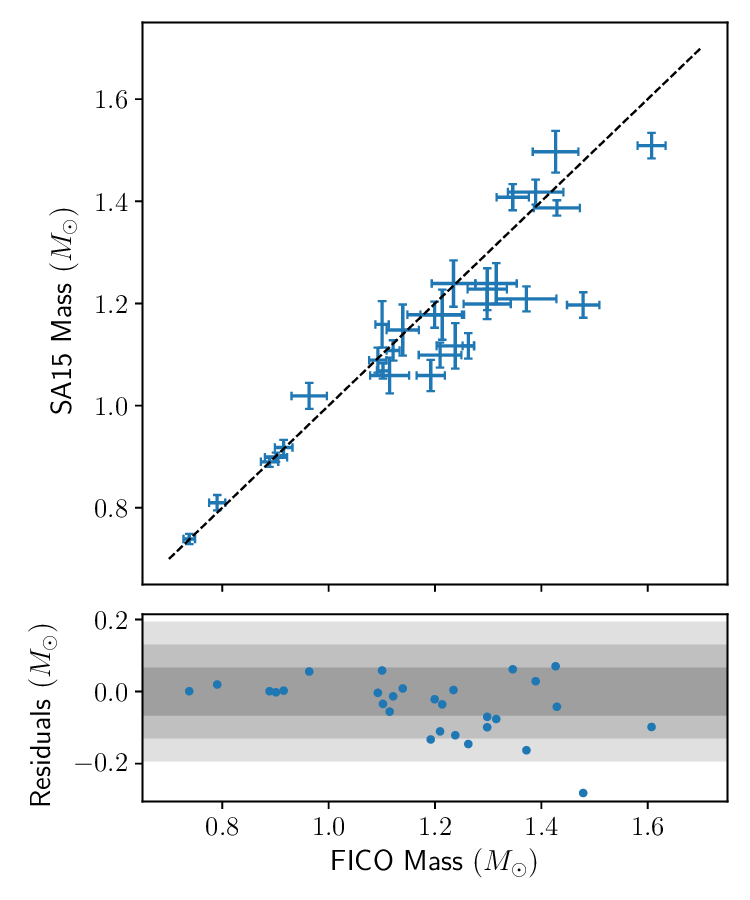}  
\end{subfigure}
\begin{subfigure}[b]{.32\textwidth}
  \includegraphics[width=.99\textwidth]{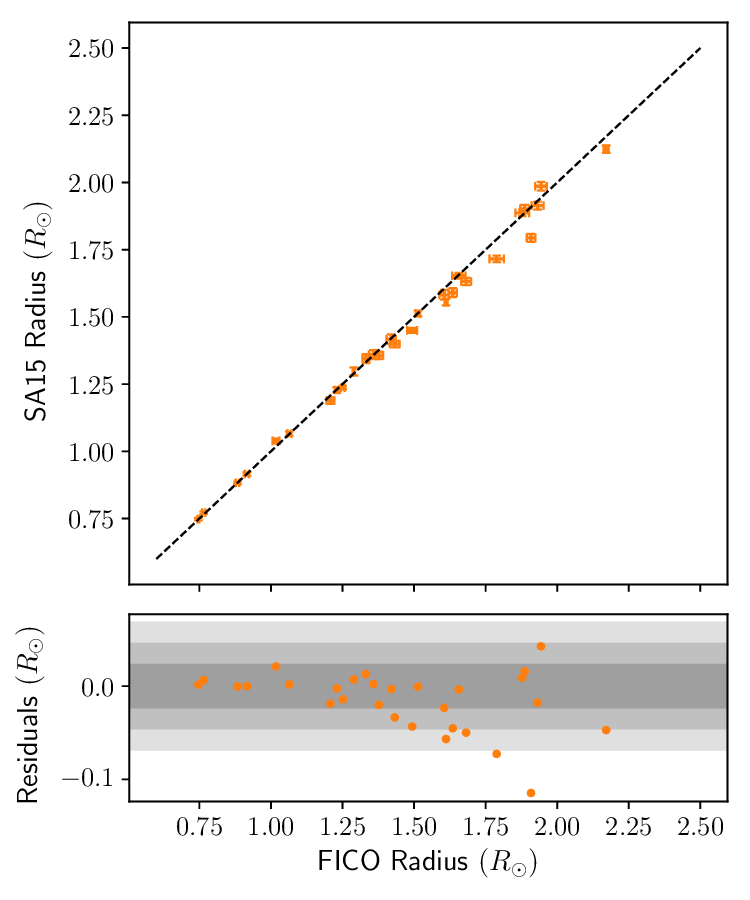}  
\end{subfigure}
\begin{subfigure}[b]{.32\textwidth}
  \includegraphics[width=.99\textwidth]{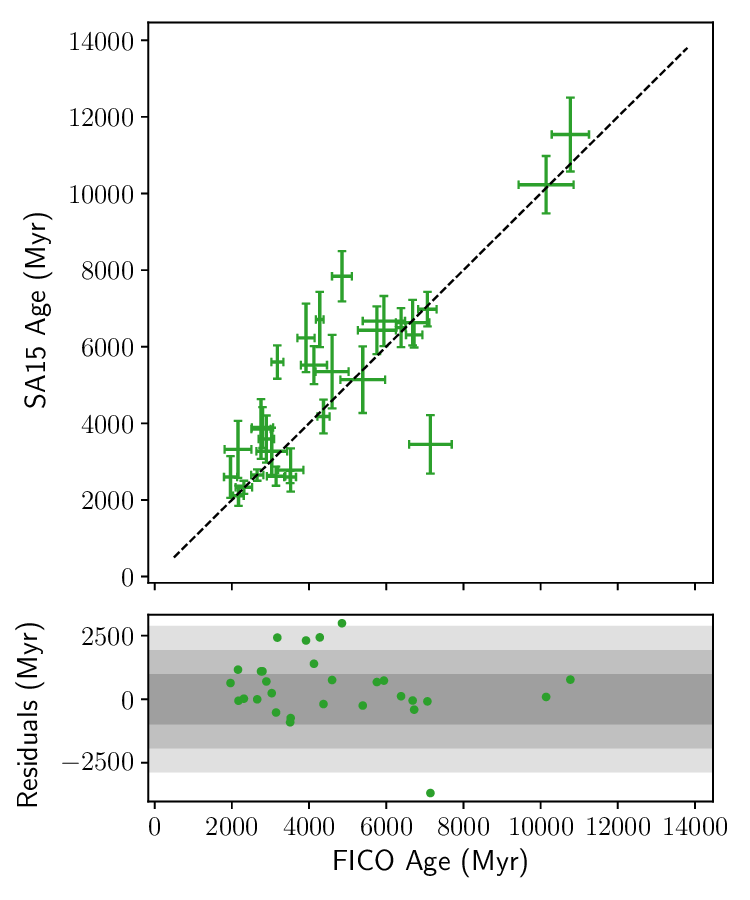}  
\end{subfigure}
\caption{Comparison between the FICO results and those of \citet{Silva-Aguirre2015} for the \citet{Davies2016} sample. The upper panel presents a mirror plot contrasting the two sets of stellar parameters, while the lower panel illustrates the corresponding residuals. Shaded regions indicate the 1$\sigma$ (dark grey), 2$\sigma$ (medium grey), and 3$\sigma$ (light grey) intervals of the residual distributions. \textit{From left to right:} stellar mass, radius, and age.}
\label{fig:comparision_with_SA15}
\end{figure*}

Figure~\ref{fig:comparision_with_SA15} presents a comparative analysis between the FICO results and those reported by \citet{Silva-Aguirre2015} for the stellar sample of \citet{Davies2016}, focusing on mass (left panel), radius (centre), and age (right). The upper panels display mirrored plots contrasting the two sets of stellar parameters, while the lower panels depict the corresponding residuals. A positive residual indicates that \citet{Silva-Aguirre2015} derived a larger value.

One star, KIC4349452, stands out distinctly in the comparative plot. Furthermore, an examination of the normalised residuals, quantifying the deviation of our results from those of \citet{Silva-Aguirre2015} in units of their combined uncertainties, revealed that KIC8554498, KIC10514430, and KIC11853905 also show significant discrepancies, exceeding the $3\sigma$ threshold. As discussed in the main text, these differences are most likely attributable to the poor quality of the seismic data available for these particular targets.

\begin{figure*}[t!]
\centering
\begin{subfigure}[b]{.4\textwidth}
  \includegraphics[width=.99\textwidth]{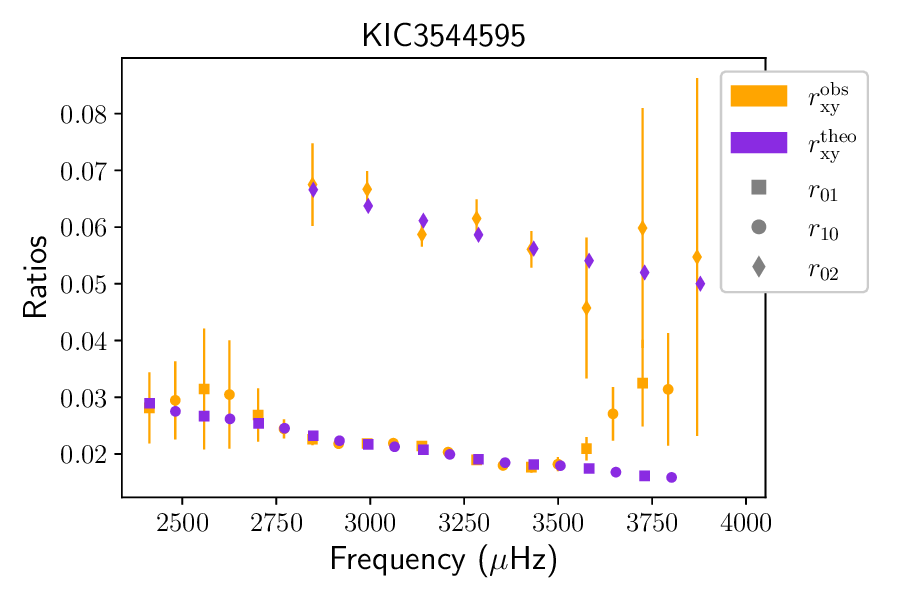}  
\end{subfigure}
\begin{subfigure}[b]{.4\textwidth}
  \includegraphics[width=.99\textwidth]{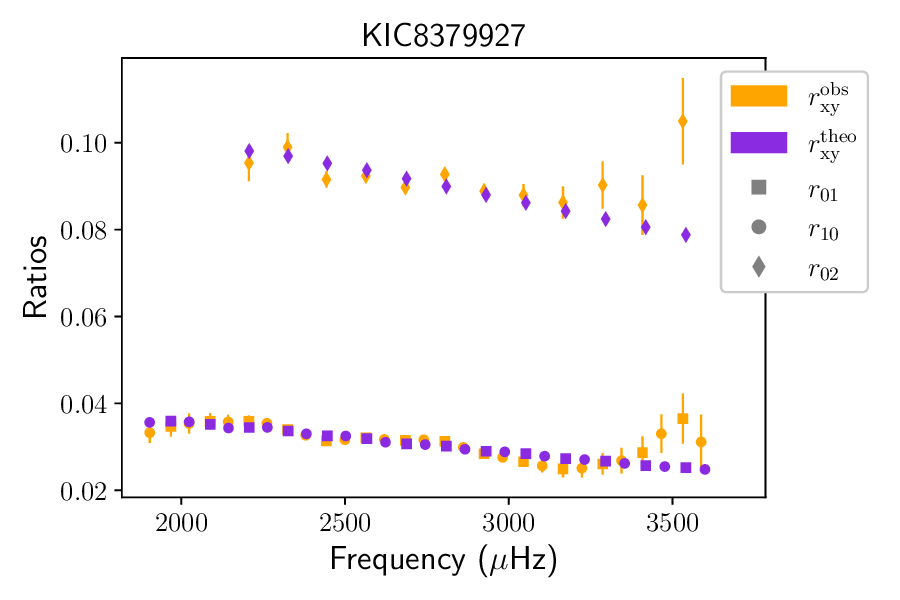}  
\end{subfigure}
\begin{subfigure}[b]{.4\textwidth}
  \includegraphics[width=.99\textwidth]{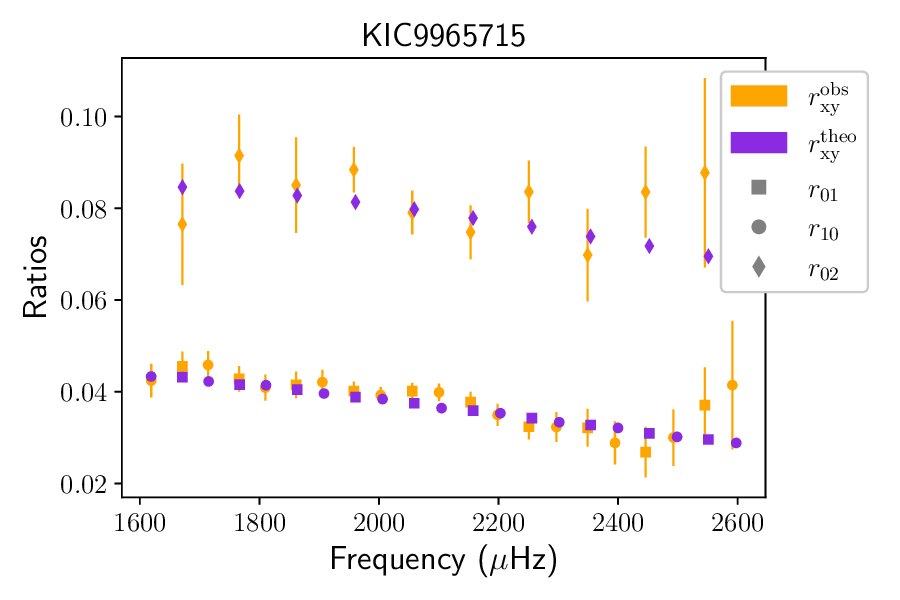}  
\end{subfigure}
\begin{subfigure}[b]{.4\textwidth}
  \includegraphics[width=.99\textwidth]{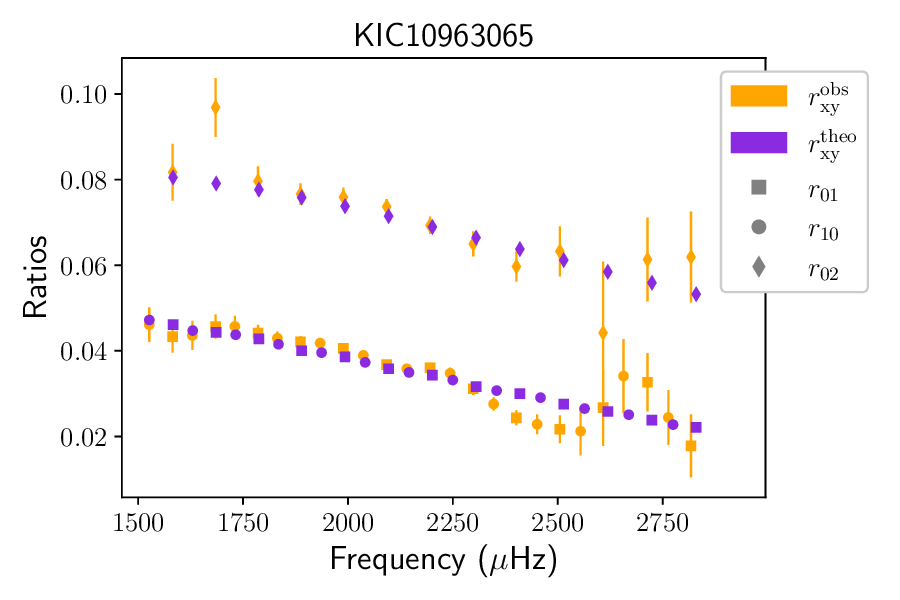}  
\end{subfigure}
\caption{Illustrative cases of high-frequency discrepancies in $r_{01}$ and $r_{10}$ ratios. The observed ratios are shown in orange and the best-fit model is displayed in purple.}
\label{fig:high_freq_discrepancies_r010}
\end{figure*}

\underline{\textit{High-frequency discrepancies in $r_{01}$ and $r_{10}$ ratios}}
\vspace*{-0.5ex}

Out of the 95 stellar targets analysed, 21 exhibit notable discrepancies in the $r_{01}$ and $r_{10}$ frequency separation ratios at high frequencies. Figure~\ref{fig:high_freq_discrepancies_r010} presents a selection of illustrative cases. In many of these, such as KIC 3544595 and KIC 9965715, we observe a clear excess in the highest-frequency modes. However, in datasets with better data-quality (where more oscillation modes are resolved), such as KIC 8379927 and KIC 10963065, this excess reveals a more complex structure: a bump-like feature that transitions into a dearth at the highest frequencies.

Several studies have already documented these discrepancies in some targets \citep[e.g.][]{Thomas2021,Betrisey2022,Nsamba2022}, but the origin of this signal remains uncertain. It does not resemble the typical helium ionization glitch \citep[see e.g.][and references therein]{Verma2019,Farnir2023}, nor does it match the signatures of non-standard physical processes such as turbulent diffusion or convective undershooting \citep[see e.g.][and references therein]{Buldgen2025_lithium}, phenomena not included in our stellar models. This possibly suggests that other unmodelled structural differences may be responsible. Alternatively, because the signal is present in the frequency separation ratios, it implies a differential impact across harmonic degrees. This opens the possibility that magnetic activity, known to influence modes of varying angular degree differently, especially at high frequencies, could be contributing to the observed signal \citep[see e.g.][]{Thomas2021}.

\FloatBarrier

\clearpage
\onecolumn

\section{Complete list of fundamental stellar properties}
\label{sec:app:complete_fundamental_stellar_data}
{\small
\begin{longtable}{lccccccc}
\caption{Fundamental properties of the stellar catalogue.}\\
\hline \hline
KIC ID & $M/M_\odot$ & $R/R_\odot$ & Age (Myr) & $\bar{\rho}_{\mathrm{inv}}$ (g/cm$^3$) & $L/L_\odot$ & Planets & Comments \\ 
\endfirsthead
\caption{continued.}\\
\hline \hline
KIC & $M/M_\odot$ & $R/R_\odot$ & Age (Myr) & $\bar{\rho}_{\mathrm{inv}}$ (g/cm$^3$) & $L/L_\odot$ & Planets & Comments \\
\hline
\endhead
\hline
\multicolumn{8}{p{\textwidth}}{
  \textbf{Notes.} From left to right, the columns present: the Kepler Input Catalog (KIC) identification number; fundamental stellar parameters including mass, radius, age, mean density, and absolute luminosity; the number of confirmed exoplanets; and our comments. Notably, the comments highlight cases where high-frequency discrepancies were observed in the seismic ratios $r_{01}$ and $r_{10}$ (abbreviated as $r_{010}$). Targets marked with an asterisk (*) indicate lower-quality interpolation, which likely led to an underestimation of the statistical uncertainty.
} \\
\multicolumn{8}{p{\textwidth}}{
  \textbf{References.}
  \textsuperscript{\textit{(a)}}\citet{Howell2012},
  \textsuperscript{\textit{(b)}}\citet{Marcy2014},
  \textsuperscript{\textit{(c)}}\citet{Cochran1997},
  \textsuperscript{\textit{(d)}} NASA Exoplanet Archive (\url{https://exoplanetarchive.ipac.caltech.edu/}).
} \\
\endfoot
\hline
\endlastfoot
\hline
\noalign{\vskip 0.3ex}
\multicolumn{8}{l}{\textit{\citet{Lund2017}}}\\
KIC1435467 & $1.362\pm 0.048$ & $1.701\pm 0.020$ & $2422\pm 194$ & $0.3894\pm 0.0023$ & $4.218\pm 0.118$ & - & - \\ 
KIC2837475 & $1.421\pm 0.039$ & $1.625\pm 0.016$ & $1741\pm 372$ & $0.4738\pm 0.0027$ & $4.612\pm 0.140$ & - & High-freq. excess in $r_{010}$ \\ 
KIC3427720 & $1.151\pm 0.026$ & $1.133\pm 0.009$ & $2244\pm 96$ & $1.1192\pm 0.0034$ & $1.580\pm 0.026$ & - & - \\ 
KIC3456181 & $1.342\pm 0.049$ & $2.071\pm 0.025$ & $2815\pm 191$ & $0.2135\pm 0.0011$ & $6.207\pm 0.150$ & - & - \\ 
KIC3632418 & $1.355\pm 0.025$ & $1.893\pm 0.013$ & $3644\pm 123$ & $0.2784\pm 0.0010$ & $4.601\pm 0.061$ & $1^{a}$ & - \\ 
KIC3656476$^*$ & $1.024\pm 0.003$ & $1.285\pm 0.001$ & $7000\pm 150$ & $0.6760\pm 0.0004$ & $2.021\pm 0.003$ & - & - \\ 
KIC3735871 & $1.169\pm 0.027$ & $1.121\pm 0.010$ & $1676\pm 160$ & $1.1717\pm 0.0038$ & $1.554\pm 0.029$ & - & - \\ 
KIC4914923$^*$ & $1.064\pm 0.006$ & $1.343\pm 0.004$ & $5575\pm 237$ & $0.6151\pm 0.0026$ & $2.443\pm 0.009$ & - & - \\ 
KIC5184732 & $1.185\pm 0.029$ & $1.332\pm 0.011$ & $4454\pm 143$ & $0.7058\pm 0.0042$ & $2.022\pm 0.024$ & - & - \\ 
KIC5773345 & $1.582\pm 0.026$ & $2.040\pm 0.013$ & $2137\pm 106$ & $0.2599\pm 0.0023$ & $6.145\pm 0.119$ & - & - \\ 
KIC5950854 & $0.961\pm 0.007$ & $1.232\pm 0.004$ & $8697\pm 683$ & $0.7253\pm 0.0032$ & $1.780\pm 0.021$ & - & - \\ 
KIC6106415 & $1.084\pm 0.012$ & $1.222\pm 0.005$ & $4494\pm 119$ & $0.8340\pm 0.0013$ & $1.837\pm 0.021$ & - & - \\ 
KIC6116048 & $1.059\pm 0.011$ & $1.240\pm 0.004$ & $5856\pm 187$ & $0.7812\pm 0.0014$ & $1.865\pm 0.021$ & - & High-freq. excess in $r_{010}$ \\ 
KIC6225718 & $1.232\pm 0.027$ & $1.262\pm 0.011$ & $2422\pm 83$ & $0.8591\pm 0.0033$ & $2.123\pm 0.028$ & - & - \\ 
KIC6508366 & $1.560\pm 0.039$ & $2.208\pm 0.018$ & $2707\pm 211$ & $0.2082\pm 0.0013$ & $6.728\pm 0.189$ & - & - \\ 
KIC6603624$^*$ & $1.039\pm 0.005$ & $1.161\pm 0.002$ & $8073\pm 191$ & $0.9361\pm 0.0007$ & $1.238\pm 0.010$ & - & - \\ 
KIC6679371 & $1.615\pm 0.022$ & $2.246\pm 0.009$ & $1736\pm 80$ & $0.2056\pm 0.0008$ & $7.670\pm 0.110$ & - & - \\ 
KIC6933899 & $1.156\pm 0.015$ & $1.597\pm 0.009$ & $5867\pm 229$ & $0.3999\pm 0.0025$ & $2.999\pm 0.054$ & - & - \\ 
KIC7103006 & $1.521\pm 0.046$ & $1.973\pm 0.020$ & $2055\pm 154$ & $0.2786\pm 0.0022$ & $5.751\pm 0.172$ & - & - \\ 
KIC7106245 & $0.992\pm 0.018$ & $1.131\pm 0.008$ & $5928\pm 401$ & $0.9658\pm 0.0057$ & $1.572\pm 0.032$ & - & - \\ 
KIC7206837 & $1.394\pm 0.045$ & $1.586\pm 0.017$ & $1929\pm 181$ & $0.4938\pm 0.0025$ & $3.823\pm 0.093$ &  - & - \\ 
KIC7296438 & $1.150\pm 0.019$ & $1.391\pm 0.010$ & $6929\pm 178$ & $0.6087\pm 0.0050$ & $2.110\pm 0.032$ & - & - \\ 
KIC7510397 & $1.318\pm 0.033$ & $1.841\pm 0.017$ & $3624\pm 145$ & $0.2950\pm 0.0012$ & $4.497\pm 0.072$ & - & - \\ 
KIC7680114 & $1.109\pm 0.011$ & $1.410\pm 0.004$ & $6821\pm 301$ & $0.5599\pm 0.0032$ & $2.313\pm 0.024$ & - & - \\ 
KIC7771282 & $1.215\pm 0.046$ & $1.620\pm 0.020$ & $3533\pm 295$ & $0.4046\pm 0.0019$ & $3.531\pm 0.093$ & - & - \\ 
KIC7871531 & $0.828\pm 0.013$ & $0.868\pm 0.005$ & $9064\pm 604$ & $1.7755\pm 0.0060$ & $0.714\pm 0.021$ & - & High-freq. excess in $r_{010}$ \\ 
KIC7940546 & $1.340\pm 0.026$ & $1.926\pm 0.014$ & $3547\pm 117$ & $0.2622\pm 0.0010$ & $4.959\pm 0.061$ & - & - \\ 
KIC7970740 & $0.739\pm 0.009$ & $0.766\pm 0.004$ & $10258\pm 315$ & $2.3171\pm 0.0138$ & $0.407\pm 0.013$ & - & High-freq. excess in $r_{010}$ \\ 
KIC8006161 & $1.006\pm 0.019$ & $0.936\pm 0.006$ & $5503\pm 233$ & $1.7266\pm 0.0009$ & $0.600\pm 0.016$ & - & - \\ 
KIC8150065 & $1.238\pm 0.047$ & $1.416\pm 0.018$ & $3757\pm 351$ & $0.6160\pm 0.0037$ & $2.657\pm 0.094$ & - & - \\ 
KIC8179536 & $1.294\pm 0.041$ & $1.365\pm 0.016$ & $2374\pm 242$ & $0.7193\pm 0.0037$ & $2.745\pm 0.061$ & - & High-freq. excess in $r_{010}$ \\ 
KIC8228742 & $1.305\pm 0.021$ & $1.840\pm 0.011$ & $4030\pm 127$ & $0.2920\pm 0.0009$ & $4.354\pm 0.047$ & - & - \\ 
KIC8379927 & $1.145\pm 0.024$ & $1.128\pm 0.009$ & $1590\pm 78$ & $1.1197\pm 0.0045$ & $1.700\pm 0.028$ & - & High-freq. excess in $r_{010}$ \\ 
KIC8394589 & $1.087\pm 0.017$ & $1.180\pm 0.007$ & $3812\pm 218$ & $0.9317\pm 0.0055$ & $1.874\pm 0.033$ & - & High-freq. excess in $r_{010}$ \\ 
KIC8424992 & $0.931\pm 0.016$ & $1.052\pm 0.005$ & $9976\pm 757$ & $1.1275\pm 0.0067$ & $1.086\pm 0.024$ & - & - \\ 
KIC8694723$^*$ & $1.085\pm 0.007$ & $1.524\pm 0.004$ & $5098\pm 290$ & $0.4242\pm 0.0014$ & $3.098\pm 0.055$ & - & - \\ 
KIC8760414$^*$ & $0.792\pm 0.007$ & $1.017\pm 0.005$ & $11450\pm 144$ & $1.0620\pm 0.0061$ & $1.163\pm 0.010$ & - & - \\ 
KIC9025370 & $0.979\pm 0.016$ & $1.002\pm 0.006$ & $4687\pm 138$ & $1.3665\pm 0.0037$ & $1.142\pm 0.022$ & - & - \\ 
KIC9098294 & $0.984\pm 0.012$ & $1.148\pm 0.006$ & $8084\pm 247$ & $0.9154\pm 0.0054$ & $1.404\pm 0.019$ & - & - \\ 
KIC9139151 & $1.201\pm 0.031$ & $1.167\pm 0.011$ & $2122\pm 145$ & $1.0649\pm 0.0063$ & $1.651\pm 0.039$ & - & - \\ 
KIC9139163 & $1.424\pm 0.039$ & $1.569\pm 0.015$ & $2007\pm 151$ & $0.5197\pm 0.0031$ & $3.654\pm 0.091$ & - & - \\ 
KIC9206432 & $1.440\pm 0.031$ & $1.525\pm 0.011$ & $1014\pm 126$ & $0.5821\pm 0.0032$ & $3.847\pm 0.088$ & - & - \\ 
KIC9353712 & $1.509\pm 0.051$ & $2.176\pm 0.026$ & $3188\pm 197$ & $0.2065\pm 0.0012$ & $6.005\pm 0.199$ & - & - \\ 
KIC9410862 & $0.988\pm 0.016$ & $1.159\pm 0.007$ & $6518\pm 417$ & $0.8946\pm 0.0053$ & $1.662\pm 0.036$ & - & High-freq. excess in $r_{010}$ \\ 
KIC9414417 & $1.440\pm 0.040$ & $1.941\pm 0.020$ & $2915\pm 160$ & $0.2759\pm 0.0012$ & $5.117\pm 0.092$ & - & - \\ 
KIC9812850 & $1.360\pm 0.043$ & $1.796\pm 0.019$ & $3310\pm 244$ & $0.3312\pm 0.0020$ & $4.499\pm 0.160$ & - & - \\ 
KIC9955598 & $0.887\pm 0.015$ & $0.883\pm 0.006$ & $6696\pm 312$ & $1.8200\pm 0.0109$ & $0.608\pm 0.020$ & $1^{b}$ & - \\ 
KIC9965715 & $1.147\pm 0.023$ & $1.297\pm 0.009$ & $2936\pm 225$ & $0.7392\pm 0.0035$ & $2.536\pm 0.042$ & - & High-freq. excess in $r_{010}$ \\ 
KIC10068307 & $1.332\pm 0.049$ & $2.045\pm 0.027$ & $4104\pm 160$ & $0.2192\pm 0.0009$ & $4.967\pm 0.081$ & - & - \\ 
KIC10079226 & $1.215\pm 0.035$ & $1.178\pm 0.012$ & $2602\pm 299$ & $1.0451\pm 0.0033$ & $1.579\pm 0.038$ & - & - \\ 
KIC10162436 & $1.461\pm 0.026$ & $2.062\pm 0.014$ & $3651\pm 130$ & $0.2333\pm 0.0009$ & $5.127\pm 0.074$ & - & - \\ 
KIC10454113 & $1.290\pm 0.026$ & $1.286\pm 0.010$ & $1667\pm 121$ & $0.8497\pm 0.0035$ & $2.297\pm 0.033$ & - & High-freq. excess in $r_{010}$ \\ 
KIC10516096 & $1.097\pm 0.008$ & $1.410\pm 0.003$ & $6167\pm 193$ & $0.5507\pm 0.0010$ & $2.499\pm 0.015$ & - & - \\ 
KIC10644253 & $1.223\pm 0.030$ & $1.137\pm 0.010$ & $1050\pm 123$ & $1.1744\pm 0.0044$ & $1.566\pm 0.033$ & - & - \\ 
KIC10730618 & $1.358\pm 0.049$ & $1.764\pm 0.020$ & $2278\pm 189$ & $0.3496\pm 0.0018$ & $4.838\pm 0.164$ & - & - \\ 
KIC10963065 & $1.069\pm 0.017$ & $1.224\pm 0.007$ & $4198\pm 177$ & $0.8208\pm 0.0049$ & $1.957\pm 0.034$ & $1^{b}$ & - \\ 
KIC11081729 & $1.259\pm 0.049$ & $1.403\pm 0.018$ & $2032\pm 297$ & $0.6446\pm 0.0035$ & $3.173\pm 0.087$ & - & - \\ 
KIC11253226 & $1.426\pm 0.035$ & $1.615\pm 0.013$ & $1847\pm 311$ & $0.4873\pm 0.0029$ & $4.628\pm 0.091$ & - & - \\ 
\pagebreak
\noalign{\vskip 0.3ex}
\multicolumn{8}{l}{\textit{\citet{Lund2017}}}\\
KIC11772920 & $0.812\pm 0.009$ & $0.839\pm 0.004$ & $9402\pm 918$ & $1.9386\pm 0.0115$ & $0.619\pm 0.027$ & - & - \\ 
KIC12009504 & $1.204\pm 0.025$ & $1.414\pm 0.011$ & $3829\pm 179$ & $0.5983\pm 0.0020$ & $2.642\pm 0.035$ & - & - \\ 
KIC12069127 & $1.571\pm 0.046$ & $2.285\pm 0.023$ & $2121\pm 127$ & $0.1868\pm 0.0009$ & $7.525\pm 0.161$ & - & - \\ 
KIC12069424 & $1.069\pm 0.008$ & $1.224\pm 0.005$ & $7038\pm 351$ & $0.8211\pm 0.0049$ & $1.539\pm 0.025$ & - & High-freq. excess in $r_{010}$ \\ 
KIC12069449$^*$ & $1.032\pm 0.006$ & $1.113\pm 0.002$ & $7140\pm 201$ & $1.0551\pm 0.0057$ & $1.251\pm 0.010$ & $1^{c}$ & High-freq. excess in $r_{010}$ \\ 
KIC12258514 & $1.181\pm 0.018$ & $1.569\pm 0.009$ & $4451\pm 157$ & $0.4273\pm 0.0009$ & $3.022\pm 0.024$ & - & - \\ 
KIC12317678 & $1.327\pm 0.032$ & $1.802\pm 0.014$ & $2670\pm 188$ & $0.3251\pm 0.0020$ & $5.341\pm 0.098$ & - & High-freq. excess in $r_{010}$ \\ 
\hline
\noalign{\vskip 0.3ex}
\multicolumn{8}{l}{\textit{\citet{Davies2016}}}\\
KIC3425851 & $1.214\pm 0.041$ & $1.358\pm 0.015$ & $2159\pm 345$ & $0.6862\pm 0.0041$ & $2.805\pm 0.120$ & - & High-freq. excess in $r_{010}$ \\ 
KIC3544595 & $0.901\pm 0.021$ & $0.916\pm 0.008$ & $6684\pm 435$ & $1.6504\pm 0.0042$ & $0.773\pm 0.020$ & $2^{d}$ & High-freq. excess in $r_{010}$ \\ 
KIC3632418 & $1.346\pm 0.030$ & $1.886\pm 0.015$ & $3510\pm 153$ & $0.2831\pm 0.0017$ & $4.740\pm 0.101$ & $1^{d}$ & - \\ 
KIC4141376 & $0.963\pm 0.033$ & $1.018\pm 0.012$ & $3033\pm 397$ & $1.2895\pm 0.0077$ & $1.345\pm 0.038$ & $1^{d}$ & - \\ 
KIC4349452 & $1.101\pm 0.013$ & $1.290\pm 0.004$ & $7144\pm 553$ & $0.7245\pm 0.0043$ & $1.806\pm 0.023$ & $3^{d}$ & High-freq. excess in $r_{010}$ \\ 
KIC4914423 & $1.210\pm 0.040$ & $1.493\pm 0.017$ & $5940\pm 547$ & $0.5114\pm 0.0040$ & $2.452\pm 0.080$ & $2^{d}$ & - \\ 
KIC5866724 & $1.298\pm 0.044$ & $1.433\pm 0.017$ & $2794\pm 274$ & $0.6222\pm 0.0037$ & $2.689\pm 0.079$ & $4^{d}$ & High-freq. excess in $r_{010}$ \\ 
KIC6196457 & $1.372\pm 0.056$ & $1.789\pm 0.025$ & $4127\pm 339$ & $0.3379\pm 0.0020$ & $3.661\pm 0.124$ & $3^{d}$ & - \\ 
KIC6278762 & $0.738\pm 0.011$ & $0.747\pm 0.004$ & $10771\pm 482$ & $2.4930\pm 0.0022$ & $0.387\pm 0.016$ & $5^{d}$ & - \\ 
KIC6521045$^*$ & $1.121\pm 0.012$ & $1.514\pm 0.008$ & $6384\pm 117$ & $0.4579\pm 0.0036$ & $2.641\pm 0.040$ & $4^{d}$ & - \\ 
KIC7670943 & $1.235\pm 0.041$ & $1.420\pm 0.016$ & $3524\pm 328$ & $0.6064\pm 0.0026$ & $2.752\pm 0.071$ & $1^{d}$ & - \\ 
KIC8077137 & $1.238\pm 0.035$ & $1.682\pm 0.017$ & $3922\pm 224$ & $0.3671\pm 0.0018$ & $3.646\pm 0.073$ & $2^{d}$ & - \\ 
KIC8292840 & $1.140\pm 0.030$ & $1.332\pm 0.012$ & $2755\pm 250$ & $0.6789\pm 0.0040$ & $2.641\pm 0.073$ & $3^{d}$ & - \\ 
KIC8478994 & $0.791\pm 0.015$ & $0.765\pm 0.005$ & $4597\pm 428$ & $2.4853\pm 0.0148$ & $0.494\pm 0.018$ & $4^{d}$ & High-freq. excess in $r_{010}$ \\ 
KIC8494142 & $1.389\pm 0.052$ & $1.878\pm 0.024$ & $3146\pm 237$ & $0.2960\pm 0.0016$ & $4.441\pm 0.128$ & $2^{d}$ & - \\ 
KIC8554498 & $1.479\pm 0.030$ & $1.909\pm 0.015$ & $3178\pm 156$ & $0.2975\pm 0.0012$ & $4.582\pm 0.072$ & $1^{d}$ & - \\ 
KIC8866102 & $1.298\pm 0.037$ & $1.377\pm 0.014$ & $1965\pm 169$ & $0.6946\pm 0.0033$ & $2.897\pm 0.051$ & $1^{d}$ & - \\ 
KIC9414417 & $1.429\pm 0.043$ & $1.932\pm 0.021$ & $2656\pm 158$ & $0.2787\pm 0.0013$ & $5.188\pm 0.105$ & - & - \\ 
KIC9592705 & $1.607\pm 0.026$ & $2.171\pm 0.009$ & $2311\pm 213$ & $0.2228\pm 0.0012$ & $6.079\pm 0.121$ & $1^{d}$ & - \\ 
KIC9955598 & $0.889\pm 0.016$ & $0.883\pm 0.006$ & $7067\pm 240$ & $1.8197\pm 0.0108$ & $0.608\pm 0.018$ & $1^{d}$ & - \\
KIC10514430 & $1.192\pm 0.027$ & $1.635\pm 0.014$ & $4852\pm 257$ & $0.3786\pm 0.0019$ & $3.455\pm 0.070$ & - & High-freq. excess in $r_{010}$ \\ 
KIC10586004 & $1.199\pm 0.051$ & $1.657\pm 0.024$ & $5757\pm 493$ & $0.3717\pm 0.0030$ & $2.843\pm 0.123$ & $3^{d}$ & - \\ 
KIC10666592 & $1.427\pm 0.043$ & $1.943\pm 0.020$ & $2172\pm 137$ & $0.2739\pm 0.0013$ & $5.642\pm 0.112$ & $1^{d}$ & - \\ 
KIC10963065 & $1.093\pm 0.016$ & $1.230\pm 0.007$ & $4373\pm 158$ & $0.8207\pm 0.0023$ & $2.003\pm 0.025$ & $1^{d}$ & High-freq. excess in $r_{010}$ \\ 
KIC11133306 & $1.115\pm 0.037$ & $1.208\pm 0.014$ & $5392\pm 579$ & $0.8918\pm 0.0053$ & $1.637\pm 0.051$ & $1^{d}$ & - \\ 
KIC11295426 & $1.102\pm 0.011$ & $1.251\pm 0.006$ & $6723\pm 212$ & $0.7928\pm 0.0047$ & $1.679\pm 0.028$ & $4^{d}$ & High-freq. excess in $r_{010}$ \\ 
KIC11807274 & $1.315\pm 0.039$ & $1.605\pm 0.017$ & $2894\pm 202$ & $0.4447\pm 0.0022$ & $3.600\pm 0.081$ & $2^{d}$ & - \\ 
KIC11853905$^*$ & $1.263\pm 0.011$ & $1.612\pm 0.005$ & $4279\pm 100$ & $0.4201\pm 0.0007$ & $3.074\pm 0.015$ & $1^{d}$ & - \\ 
KIC11904151 & $0.915\pm 0.017$ & $1.064\pm 0.007$ & $10142\pm 713$ & $1.0708\pm 0.0064$ & $1.104\pm 0.023$ & $3^{d}$ & -
\label{tab:complete_fundamental_stellar_data}
\end{longtable}
}

\end{document}